\documentclass[12pt]{article}

\usepackage[dvips]{graphicx}
\font\scaps=cmcsc10    

\newcommand \be{\begin{equation}}
\newcommand \ee{\end{equation}}
\newcommand \ba{\begin{eqnarray}}
\newcommand \ea{\end{eqnarray}}
\begin{document}

\def\today{\ifcase\month\or
 January\or February\or March\or April\or May\or June\or
 July\or August\or September\or October\or November\or December\fi
 \space\number\day, \number\year}
%
\hfil PostScript file created: \today{}; \ time \the\time \ minutes
\vskip .25in

%


\centerline {STATISTICAL DISTRIBUTIONS OF EARTHQUAKE NUMBERS:
}
\centerline {CONSEQUENCE OF BRANCHING PROCESS
}

\vskip .25in
\begin{center}
{Yan Y. Kagan, }
\end{center}
\centerline {Department of Earth and Space Sciences,
University of California,}
\centerline {Los Angeles, California 90095-1567, USA;}
\centerline {Emails: {\tt ykagan@ucla.edu,
kagan@moho.ess.ucla.edu }}
\vskip 0.02 truein

\vspace{0.25cm}

\noindent
{\bf Abstract.}

We discuss various statistical distributions of earthquake
numbers.
Previously we derived several discrete distributions to
describe earthquake numbers for the branching model of
earthquake occurrence: these distributions are the Poisson,
geometric, logarithmic, and the negative binomial (NBD).
The theoretical model is the `birth and immigration'
population process.
The first three distributions above can be considered
special cases of the NBD.
In particular, a point branching process along the magnitude
(or log seismic moment) axis with independent events
(immigrants) explains the magnitude/moment-frequency
relation and the NBD of earthquake counts in large time/space
windows, as well as the dependence of the NBD parameters on
the magnitude threshold (magnitude
of an earthquake catalog completeness).
We discuss applying these distributions, especially the NBD,
to approximate event numbers in earthquake catalogs.
There are many different representations of the NBD.
Most can be traced either to the Pascal distribution or to the
mixture of the Poisson distribution with the gamma law.
We discuss advantages and drawbacks of both representations
for statistical analysis of earthquake catalogs.
We also consider applying the NBD to earthquake forecasts and
describe the limits of the application for the given
equations.
In contrast to the one-parameter Poisson distribution so
widely used to describe earthquake occurrence, the NBD
has two parameters.
The second parameter can be used to characterize clustering or
over-dispersion of a process.
We determine the parameter values and their uncertainties for
several local and global catalogs, and their subdivisions in
various time intervals, magnitude thresholds, spatial windows,
and tectonic categories.
The theoretical model of how the clustering parameter depends
on the corner (maximum) magnitude can be used to predict
future earthquake number distribution in regions where very
large earthquakes have not yet occurred.

\vskip .15in
\noindent
{\bf Short running title}:
{\sc
Distributions of earthquake numbers
}



\section{Introduction}
\label{intro}

Earthquake forecasts are important in estimating hazard and
risk and in making informed decisions to manage emergency
response.
How can we establish standards for reporting and testing
earthquake forecasts?
One significant effort began in California, where the
Regional Earthquake Likelihood Models (RELM) project published
a dozen models for earthquake rate density and a likelihood
based method for prospective testing (Field, 2007; Schorlemmer
and Gerstenberger, 2007; Schorlemmer {\it et al.}, 2007;
Schorlemmer {\it et al.}, 2009).
The Collaboratory for Study of Earthquake Predictability
(CSEP) is currently extending the tests to several natural
laboratories around the globe.

One standard test is to compare the number of predicted
earthquakes with the actual number of events during the test
period.
To do so we need to know the statistical distribution of
earthquake numbers.

The standard assumption long used in earthquake hazard
analysis (Cornell, 1968) is that earthquake occurrence
is reasonably well described by the Poisson distribution.
However, it has also been known for a long time that
earthquakes are clustered in time and space: their
distribution is over-dispersed compared to the Poisson law.
One conventional way to treat this problem is to decluster an
earthquake catalog (Schorlemmer {\it et al.}, 2007).
But there are several declustering procedures, mostly based on
{\sl ad-hoc} rules.
Hence declustered catalogs are not unique and usually not
fully reproducible.
Therefore, it is important to derive and investigate
earthquake number distribution in real earthquake catalogs.

Kagan (1996), Jackson and Kagan (1999), and Kagan and Jackson
(2000) have all used the negative binomial distribution (NBD)
to approximate earthquake numbers in catalogs.
The NBD has a higher variance than the Poisson law and can
be shown (Kagan, 1973a,b) to result from the branching nature
of earthquake occurrence.

In principle, several other over-dispersed discrete
distributions, such as generalized Poisson distributions
(Consul, 1989) or generalized negative-binomial distributions
(Tripathi, 2006; Hilbe, 2007) can be used to approximate
earthquake numbers.
However, the NBD has the advantage of relative simplicity and
is supported by theoretical arguments (Kagan, 1973a,b).
As we discuss below, in addition to negative-binomial and
Poisson distributions, several other statistical discrete
distributions can describe earthquake numbers.
A general discussion of such distributions can be found in
Johnson {\it et al.}\ (2005) and Kotz {\it et al.}\ (2006).

Over the years many papers have analyzed various aspects of
earthquake numbers distributions; for example, see Vere-Jones
(1970), Shlien and Toks\"oz (1970), Dionysiou and Papadopoulos
(1992).
These publications investigated the distributions empirically
by counting earthquake numbers in catalogs and trying to
approximate them by various statistical laws.
Here we explain these distributions as a consequence of the
stochastic branching model.

Therefore, in addition to the NBD and the Poisson
distributions, in this study we will investigate the geometric
and logarithmic distributions in several earthquake catalogs
and show their applicability in certain conditions.
After presenting the theoretical derivation of these
distributions, we explore the statistical parameter estimation
for these laws.
Then we apply these methods to several earthquake catalogs.
Two global (CMT and PDE) and one local Californian catalog
are studied and the results are displayed in tables and
diagrams.
These results can be used in earthquake forecast testing
(Schorlemmer {\it et al.}, 2007; Schorlemmer {\it et al.},
2009; Kagan {\it et al.}, 2009).

\section {Theoretical considerations
}
\label{theo}

\subsection {Generating function for the NBD
}
\label{gen1}

Several stochastic models of earthquake occurrence were
proposed and almost all were based on the theory of branching
processes (Kagan, 2006).
Branching is expected to model the well-known property of
primary and secondary clustering for aftershock sequences: a
strong aftershock (or foreshock) tends to have its own
sequence of dependent events.
These multidimensional models are

\begin{figure}
\begin{center}
\includegraphics[width=0.75\textwidth]{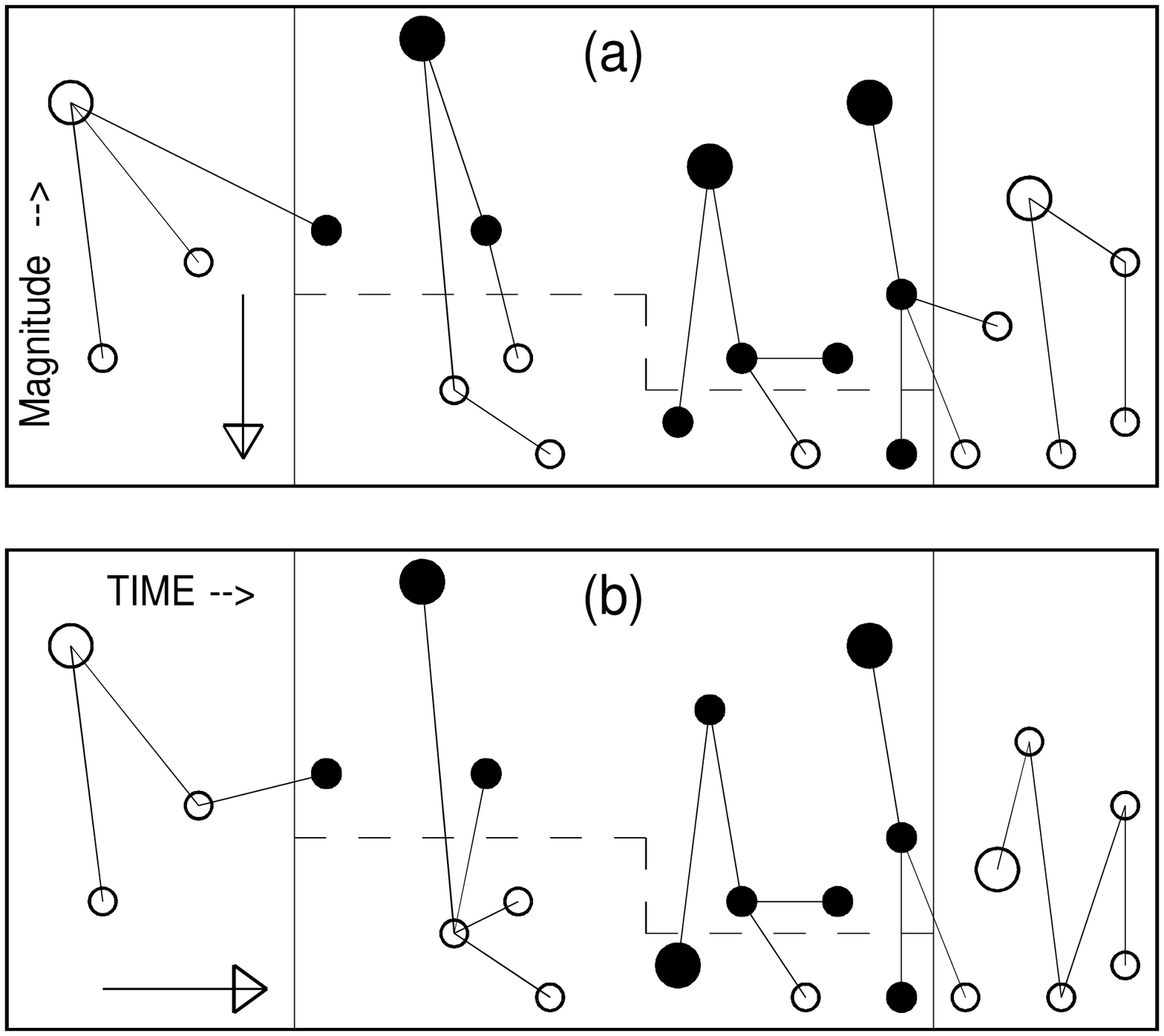}
\caption{\label{fig01}
}
\end{center}
Earthquake branching models:
filled circles indicate observed earthquakes; open circles
are unobserved, modeled events.
Vertical thin lines separate unknown past events, the current
catalog, and future events.
The dashed line represents a magnitude threshold; the
earthquake record above the threshold is considered to be
complete.
Many small events are not registered below this threshold,
though some events are observed even below the threshold.
Large circles denote the initial (or main) event of a
cluster.
Arrows indicate the direction of the branching process:
down magnitude axis in (a) and up time axis in
(b).
\hfil\break
(a) Branching-in-moment (magnitude) model.
(b) Branching-in-time model.
\end{figure}

$\bullet$
(a) The supercritical point process branching along the
magnitude axis, introduced by Kagan (1973a,b) and shown in
Fig.~\ref{fig01}a.
Earthquake occurrence constitutes a down-going cascade in this
model.

$\bullet$
(b) Critical (or rather slightly subcritical) point process
branching along the time axis (Hawkes 1971; Kagan and Knopoff,
1987; Ogata, 1988)~-- often called the Hawkes self-exciting
process or the ETAS model (see Fig.~\ref{fig01}b).
Hawkes and Tukey debate (see discussion section in Kagan,
1973b) the difference between branching in earthquake size and
that in time.
Bremaud and Massoulie (2001) recently proposed a variant of
Hawkes' process with no independent events (immigrants).
However, in earthquake catalogs limited in time-span, we need
to introduce independent events.
Otherwise, models would be highly non-unique.

Both models shown in Fig.~\ref{fig01} use the Poisson cluster
process to approximate earthquake occurrence.
Earthquake clusters are assumed to follow the Poisson
occurrence.
Earthquakes within a cluster are modeled by a multidimensional
branching process, which reproduces a temporal-spatial pattern
of dependent events (mostly aftershocks) around the initial
one in a sequence (Kagan, 1973a,b; Kagan and Knopoff, 1987;
Ogata, 1988, 1998).

In one form or another, these models employ the classical
statistical properties of earthquake occurrence: the
Gutenberg-Richter (G-R) relation and Omori's law.
Model (a) reproduces the G-R relation as the result of
a supercritical branching along the magnitude axis, while the
temporal distribution (the Omori-type law) must be imposed.
In model (b) both the temporal and magnitude distributions
are imposed.

The simplest way to describe discrete distributions is by
using the probability generating function (Bartlett, 1978;
Evans {\it et al.}, 2000).
Given the generating function $\phi \, (z)$, the probability
function $f \, (k)$ can be obtained as
\be
f \, (k) \ = \ {1 \over {k!} } \left [
{ { d^k \phi \, (z) } \over { d z^k} }
\right ]_{z=0}
\, .
\label{NBD_Eq01}
\ee

Following the graph of inter-earthquake connections as shown
in Fig.~\ref{fig01}a, we investigate the earthquake numbers
distributions for space-time intervals larger that the average
dimensions of earthquake clusters.
Thus, we neglect space-time differences between cluster
members.
We assume that the independent (mainshocks) and dependent
(foreshocks-aftershocks) events are occurring along the $\Xi
\, = \, \log M$ axis ($M$ is the scalar seismic moment) with a
constant rates
\ba
\nu \, (d \, \Xi ) \ &=& \ \nu \, \cdot \, d \, \Xi \ = \
{\rm const} \, \cdot \, d \, \Xi \, ,
\nonumber\\
\beta \, (d \, \Xi ) \ &=& \ \beta \, \cdot \, d \, \Xi \ =
\ {\rm const} \, \cdot \, d \, \Xi \, ,
\nonumber\\
&& {\rm for} \
\Xi \, < \, \Xi_{\rm m} \, ,
\label{NBD_Eq02}
\ea
where $\Xi_{\rm m}$ is the logarithm of the maximum seismic
moment, $M_m$; $\nu$ and $\beta $ are rates of
independent and dependent events, respectively.

Events occurrence can be modeled as an `immigration and
birth' process, where independent, spontaneous
earthquakes (mainshocks) are treated as `immigrants'.
Any immigrant may spawn offspring, who may spawn their own
offspring, etc., with the whole family making up a cluster.
Seismologists generally call the largest family member the
`mainshock', and the preceding smaller events the
`foreshocks', and subsequent members the
`aftershocks'.
In this case the conditional generating function for the
number of events with $\Xi \ge \Xi_c$ in a cluster, including
the mainshock, is (Bartlett, 1978, Eq.~3.4(7), p.~76)
\be
\phi \, (z \, | \, \Xi_c) \ = \ { {z \, (M_c/M)^\beta }
\over {1 \, - \, z \, [\, 1 \, - \, (M_c/M)^\beta \, ]} }
\, ,
\label{NBD_Eq03}
\ee
where $\beta$ is the index of the seismic moment distribution,
$\beta = {2 \over 3} b$; \ $b$ is the parameter of the
G-R law (Kagan and Jackson, 2000; Bird and Kagan, 2004), and
$M_c$ is the moment detection threshold of a seismographic
network ($M \ge M_c$).
In this formula we modify Bartlett's equation for the `birth
and immigration' (but no `deaths') population process.

In future calculations here we take
\be
\beta \ = \ 0.62
\, :
\label{NBD_Eq04}
\ee
the value suggested by results in the Bird and Kagan (2004)
and Kagan {\it et al.}\ (2009).

Equation (\ref{NBD_Eq03}) characterizes the geometric
distribution, with the probability function
\be
f \, (k) \ = \ (1 - p)^{k-1} p \quad {\rm for} \quad k=1,2,3,
\dots
\,
\label{NBD_Eq05}
\ee
A more common form of the geometric distribution (Evans {\it
et al.}, 2000) is
\be
f \, (k) \ = \ (1 - p)^k p \quad {\rm for} \quad k=0,1,2,
\dots
\,
\label{NBD_Eq06}
\ee
It would be appropriate for the number of dependent shocks
in a cluster.
Its conditional generating function is
\be
\phi \, (z \, | \, \Xi_c) \ = \ { {(M_c/M)^\beta }
\over {1 \, - \, z \, [\, 1 \, - \, (M_c/M)^\beta \, ]} }
\, .
\label{NBD_Eq07}
\ee
The geometric distribution is a discrete analog of the
exponential distribution for continuous variables.
For a small $p$-value
\be
f \, (k) \ \approx \ p \, \exp - k \, p
\, .
\label{NBD_Eq08}
\ee

For the distribution of the total number of events in an
earthquake catalog we obtain (Bartlett, 1978, Ch.~3.41, p.~83)
\ba
\phi \, (z ) \ &=& \ \exp \int_{\Xi_c}^{\Xi_m}
[ \, \phi \, (z \, | \, \Xi) \, - \, 1 \, ] \, \nu \, d \,
\Xi
\nonumber\\
&=& \ \exp \left \{ \left [ { { \log
\left ( 1 \, - \, z \, [\, 1 \, - \, (M_c/M_m)^\beta \, ]
\right ) }
\over
- {\beta \, (\, \Xi_m - \Xi_c \, ) } } \, - 1 \, \right ] \,
\nu \, (\, \Xi_m - \Xi_c \, ) \right \}
\nonumber\\
&=& \ \left [
{ {(M_c/M_m)^\beta }
\over {1 \, - \, z \, [\, 1 \, - \, (M_c/M_m)^\beta \, ]} }
\right ]^{\nu/\beta }
\, .
\label{NBD_Eq09}
\ea
The above generating function is for the NBD.

In this derivation we assume that the `immigration and birth'
process starts at the maximum moment: we assume the
moment distribution to be a truncated Pareto distribution with
a `hard' limit on the maximum moment (Kagan and Jackson, 2000;
Zaliapin {\it at al.}, 2005; Kagan, 2006).
However, a more appropriate seismic moment distribution (a
tapered Pareto distribution) uses a `soft' limit or a corner
moment ({\it ibid.}, Bird and Kagan, 2004).
To consider this, Kagan and Jackson (2000, Eq.~29) propose
some modification of the NBD formulas.

The last line in (\ref{NBD_Eq09}) means that the distribution
of events (counting mainshocks as well) in earthquake clusters
would be logarithmic (or logarithmic series) with the
probability function
\be
f \, (k) \ = \
{ { \left [1 \, - \, (M_c/M_m)^\beta \right ]^k }
\over {k \, \log (M_c/M_m)^\beta } }
\ = \
{ { \left [1 \, - \, (M_c/M_m)^\beta \right ]^k }
\over {k \, \beta \, (\Xi_m - \Xi_c) } }
\quad {\rm for} \quad k=1,2,3,
\dots
\, .
\label{NBD_Eq10}
\ee
The standard form of the logarithmic distribution
(Evans {\it et al.}, 2000) is
\be
f \, (k) \ = \
{ { -1 } \over {\log (1-p) } } \cdot
{ { p^k } \over k }
\quad {\rm for} \quad k=1,2,3,
\dots
\, .
\label{NBD_Eq11}
\ee
For a small moment threshold, the logarithmic distribution
can be approximated by a tapered Pareto distribution (Zaliapin
{\it et al.}, 2005)
\be
f \, (k) \ \approx \ \left \{ \, k \log \left [ \,
(M_m/M_c)^\beta \, \right ]
\exp \left [ \, k \, (M_c/M_m)^\beta \, \right ] \right \}
^{-1}
\quad {\rm for} \quad k=1,2,3,
\dots
\, ,
\label{NBD_Eq12}
\ee
i.e., the clusters with small and intermediate mainshocks
follow the Pareto style heavy-tail distribution.
The exponential tail should be observed only for clusters with
large mainshocks having a magnitude close to maximum.

The average size for a group of dependent events is
\be
S \, (k) \ = \
{ { \left [\, (M_m/M_c)^\beta - 1 \right ] }
\over {\beta \, (\Xi_m - \Xi_c) } }
\, .
\label{NBD_Eq13}
\ee
It is clear from Eqs.~\ref{NBD_Eq10} and \ref{NBD_Eq13} that
the number of dependent events decreases as $M_c \to M_m$:
the largest earthquakes are distributed according to
the Poisson law.

The Poisson distribution has the probability function
of observing $k$ events as
\be
f \, (k) \ = \ { {\lambda^k \exp (-\lambda) }
\over {k! } } \, .
\label{NBD_Eq14}
\ee
For this distribution its mean and variance are equal to its
rate $\lambda$.

What are the distributions of earthquake numbers for the
branching-in-time model (see Fig.~\ref{fig01}b)?
Can we calculate these distributions similarly to model (a)?
Indeed, when the first event in a sequence of earthquakes is
the largest one, as is often the case, there is little
ambiguity in calling it a mainshock, and the other dependent
events may be called aftershocks, though some of them are
actually aftershocks of aftershocks.
In such a case there is little difference between (a) and (b)
models.
If, however, the first event in a sequence is smaller than
the subsequent events, it is typically called a foreshock.
Retrospectively, it is relatively easy to subdivide an
earthquake catalogue into fore-, main-, and aftershocks.
However, in real-time forecasting it is uncertain
whether the most recent event registered by a seismographic
network is a foreshock or a mainshock.
Although the subsequent triggered events are likely to be
smaller and would thus be called aftershocks, there is a
significant chance that some succeeding earthquakes may be
bigger than the predecessors that triggered them (Kagan,
1991).

The difficulty in model (b) is that some connections between
events are not observable.
Suppose that there is a potential foreshock-mainshock
pair: a larger earthquake is preceded by a smaller one.
However, the first event is below the magnitude threshold and
the larger earthquake is above the threshold.
Then this second event would be treated as independent
(immigrant); that is, our calculations would miss this
connection (Kagan, 1991; Sornette and Werner, 2005; Kagan,
2006).
It is possible that the number of such missed connections is
small, and the distributions are similar to those derived
above.
Nevertheless, such distribution estimates for this model would
be approximate.
For the branching-in-magnitude model (Fig.~\ref{fig01}a), the
derived expressions are exact.

The equations obtained in this subsection are based on the
theory of population processes.
In simple models of such processes (Bartlett, 1978), we
assume that individuals start to reproduce immediately after
their birth.
This is not the case for an earthquake occurrence: after
a large event, the aftershocks even large ones cannot be
observed for a period of time (Kagan, 2004).
Moreover, earthquakes are not instantaneous, point events;
and their rupture takes time.
Hence, the derived formulas would over-estimate the number of
dependent events.
However, as we will see below, during analysis of the
earthquake catalogs the theoretical models provide some
valuable insight into the quantitative behavior of
earthquake occurrence.

\subsection{NBD distribution expressions
}
\label{nbd}

There are many different (often confusing) representations of
the NBD (see several examples in Anscombe, 1950; Shenton and
Myers, 1963).
The most frequently used (we call it {\sl standard}) form of
the probability density function for the NBD
generalizes the Pascal distribution (Feller, 1968,
Eq.~VI.8.1; Hilbe, 2007, Eq.~5.19):
\ba
f \, (k) \ &=& \
{ { \tau \, (\tau + 1) \dots (\tau + k - 2) \, (\tau + k - 1)}
\over {k!} } \, \times \, \theta^\tau (1 - \theta)^k \ = \
{\tau + k - 1 \choose \tau -1} \, \times \, \theta^\tau (1 -
\theta)^k \
\nonumber\\
\ &=& \
{\tau + k - 1 \choose k} \, \times \, \theta^\tau (1 -
\theta)^k \ \ = \ \
{ {\Gamma (\tau + k) } \over {\Gamma (\tau) \, k! } } \,
\times \, \theta^\tau (1 - \theta)^k \, ,
\label{NBD_Eq15}
\ea
where $ k = 0, 1, 2, \dots$, $\Gamma$ is the gamma function,
$ 0 \le \theta \le 1$, and $\tau > 0$.
If the parameter $\tau$ is integer, then this formula (Pascal
distribution) is the probability distribution of a certain
number of failures and successes in a series of independent
and identically distributed Bernoulli trials.
For $k+\tau$ Bernoulli trials with success probability
$\theta$, the negative binomial gives the probability for $k$
failures and $\tau$ successes, with success on the last trial.
If $\tau = 1$, this equation corresponds to (\ref{NBD_Eq05}):
the geometric distribution.
Therefore, the latter distribution can be considered a special
case of the NBD.

The average of $k$ for the NBD is
\be
E(k) \ = \ m_1 \ = \ \tau \, {{1 - \theta} \over \theta},
\label{NBD_Eq16}
\ee
and its variance
\be
D(k) \ = \ m_2 \ = \ \tau \, {{1 - \theta} \over \theta^2}.
\label{NBD_Eq17}
\ee
The negative binomial distribution generally has a larger
standard deviation than the Poisson law.
Thus, it is often
called an `over-dispersed Poisson distribution' (Hilbe, 2007).
For $\theta \to 1$ and $\tau (1 - \theta) \to \lambda$
expression (\ref{NBD_Eq15}) tends to (\ref{NBD_Eq14}) (Feller,
1968, p.~281); the negative binomial distribution becomes the
Poisson distribution; the latter distribution is a
special case of the former.

Anraku and Yanagimoto (1990, Eq.~1.1) and Hilbe (2007,
Eqs.~5.9 or 5.29) propose the following distribution density
form, which they obtain as a mixture of the Poisson
distributions with the gamma distributed rate parameter
$\lambda$ (see, for example, Hilbe, 2007, Eqs.~5.1--5.9 or
7.21--7.33)
\be
f \, (k) \ = \ { {\Gamma (1/\alpha + k) } \over
{\Gamma (1/\alpha) \, k! } } \cdot
{ { \lambda^k } \over {
(\lambda + 1/\alpha)^k \,
(1 + \alpha \lambda)^{1/\alpha}
} }
\ = \
{ { \lambda^k } \over { k! } } \cdot
{ {\Gamma (1/\alpha + k) } \over
{\Gamma (1/\alpha) \, (\lambda + 1/\alpha)^k } } \cdot
{ { 1 } \over
{ (1 + \alpha \lambda)^{1/\alpha} } }
\, ,
\label{NBD_Eq18}
\ee
which converges to the geometric distribution if $ \, \alpha =
1 \, $, but to the Poisson distribution (\ref{NBD_Eq14}) when
$ \, \alpha \to 0 \, $,
\be
f \, (k) \ = \
{ { \lambda^k } \over { k! } } \cdot
1
\cdot
{ { 1 } \over { \exp(\lambda) } }
\, .
\label{NBD_Eq19}
\ee
Comparison with (\ref{NBD_Eq15}) shows that
$ \, \alpha = 1/\tau \, $
and $ \, \lambda = \tau \, (1 - \theta) / \theta \, $.
Distribution (\ref{NBD_Eq18}) is called an {\sl alternative}
form of the NBD in Wikipedia and we accept this term for
our purposes.

In addition to the above representations of the NBD, we use
Evans' (1953) expressions.
Evans provides the parameter uncertainty equations for the
estimates by the moment method.
The formula for the probability density is
\be
f \, (k) \ = \ { {\Gamma (\lambda/a + k) } \over
{\Gamma (\lambda/a) \, k! } } \times
{ { [ a/(1+a)]^k } \over {(1 + a )^{\lambda/a} } }
\, ,
\label{NBD_Eq20}
\ee
If we make $a = \alpha \, \lambda $, this equation converts to
(\ref{NBD_Eq18}).

The probability generating function (Bartlett, 1978) for the
NBD standard form is
\be
\phi \, (z) \ = \ \left [ {\theta \over {1 - z(1 - \theta)} }
\right ]^{\tau }
\, ,
\label{NBD_Eq21}
\ee

For the alternative form (\ref{NBD_Eq18})
of the NBD the generating function is
\be
\phi \, (z) \ = \ {1 \over { [ 1 + \alpha \lambda - z \alpha
\lambda ]^ {1/\alpha} } }
\, ,
\label{NBD_Eq22}
\ee

Comparing (\ref{NBD_Eq21}) with (\ref{NBD_Eq09}), it is clear
that if the moment threshold is close to $M_{m}$, the NBD
approaches the Poisson distribution for earthquake counts.
This is confirmed by extensive seismological practice (see
also Figs.~\ref{fig13} and \ref{fig14} below).

After comparing (\ref{NBD_Eq21}) with (\ref{NBD_Eq09}), we
propose the following relations between the parameters
\be
\theta \ = \ \left ( {M_c \over M_m} \right )^\beta
\, ,
\label{NBD_Eq23}
\ee
and
\be
\tau \ = \ \nu / \beta
\, .
\label{NBD_Eq24}
\ee
However, as mentioned at the end of Subsection~\ref{gen1},
such relations are valid only when larger shocks start
producing dependent events immediately after their rupture.
In earthquake catalogs there is a significant delay of
aftershock generation (Kagan, 2004).
Thus, these expressions are over-estimated.
Moreover, in all available earthquake catalogs there are
very few or no mainshocks with a size approaching the
maximum magnitude.
Therefore, we would not expect the observational
$\theta$-estimates to be close to that of (\ref{NBD_Eq23}).
However, as we will see the dependence of $\theta$ on the
magnitude threshold, the maximum magnitude, and $\beta$ is
visible in earthquake records.

\subsection{Statistical parameter estimation
}
\label{stat}

For the Poisson distribution (\ref{NBD_Eq14}) the estimate of
its parameter is the average earthquake rate per time interval
${\Delta T}$
\be
\hat \lambda \ = \ { {n \, {\Delta T}} \over T } \, ,
\label{NBD_Eq25}
\ee
where $T$ is the time-span and $n$ is total number of events
in a catalog.
The estimate of $p$ for the geometric distribution's
(\ref{NBD_Eq05}) is
\be
\hat p \ = \ {1 \over {1 + m_1} } \, ,
\label{NBD_Eq26}
\ee
where $ \, m_1 = \hat \lambda \, $ is the average (the first
moment).

For the logarithmic distribution (\ref{NBD_Eq11}) there is
no simple expression to evaluate its parameter $p$.
Patil (1962, Table~1) and Patil and Wani (1965, Table~2)
propose tables for calculating the maximum likelihood
estimate (MLE) of the parameter after determining the average
number of events.

For the standard form (\ref{NBD_Eq15}) of the NBD, we use
(\ref{NBD_Eq16}) and (\ref{NBD_Eq17}) to obtain the estimates
of the NBD parameters by the moment method
\be
\hat \theta \ = \ {{m_1} \over {m_2}} \, ,
\label{NBD_Eq27}
\ee
and
\be
\hat \tau \ = \ {{m_1^2} \over {m_2 - m_1}} \, ,
\label{NBD_Eq28}
\ee
where $m_1$ and $m_2$ are the average and variance of the
empirical number distribution.
Below we sometimes would use the term {\sl moment estimate}
for the estimate by the moment method.
For the Poisson process, $m_2 = m_1$.
Hence the estimate (\ref{NBD_Eq28}) would be unstable if the
NBD process approaches the Poisson one, and the estimate
uncertainty would be high.

For the alternative form of the NBD (\ref{NBD_Eq18}), we obtain
the following moment estimates
\be
\hat \lambda \ = \ m_1 \, ,
\label{NBD_Eq29}
\ee
and
\be
\hat \alpha \ = \ {{m_2 - m_1} \over m_1^2} \, .
\label{NBD_Eq30}
\ee
Evans' (1953) parameter $a$ is estimated similarly to
(\ref{NBD_Eq30})
\be
\hat a \ = \ {{m_2 - m_1} \over m_1} \ = \ {m_2 \over m_1} - 1
\, .
\label{NBD_Eq31}
\ee
Evans (1953, p.~203, see `{\sl Estimation by Method~1}')
derives approximate estimates of the parameters' uncertainties
\be
\sigma_\lambda \ = \ \sqrt { \lambda (a + 1)/N }
\, .
\label{NBD_Eq32}
\ee
and
\be
\sigma_a \ \approx \ \sqrt { 2 (a + 1)/N + a (a + 1) (3a +
2)/(\lambda N) }
\, .
\label{NBD_Eq33}
\ee
as well as the covariance between these two estimates
\be
{\rm Cov } (\hat \lambda, \, \hat a) \ \approx \
 a (a + 1)/N
\, .
\label{NBD_Eq34}
\ee
In these equations $N$ is the number of time intervals with
earthquake counts.

The maximum likelihood estimate (MLE) of any parameters for
the discussed distributions can be obtained by maximizing
the log-likelihood function
\be
{\ell } \ = \ \log \, \prod_{j=0}^\infty \,
[\, f \, (k_j) \, ]^{P(k_j) \, } \ =
\ \sum_{j=0}^\infty \, P(k_j) \, \log \, f \, (k_j)
\, ,
\label{NBD_Eq35}
\ee
where $P(k_j)$ is the observational frequency of earthquake
numbers in interval $j$.
Function $f(k)$ is defined by the expression
(\ref{NBD_Eq14}), (\ref{NBD_Eq15}), (\ref{NBD_Eq18}),
or (\ref{NBD_Eq20}) for the Poisson, the standard NBD, the
alternative NBD, and Evans' formula, respectively.
To evaluate parameter uncertainties we need to obtain the
Hessian matrix (the second partial derivatives of the
likelihood function) of the parameter estimates
(Wilks, 1962; Hilbe, 2007).

\section {Earthquake catalogs
}
\label{catl}

We studied earthquake distributions and clustering for the
global CMT catalog of moment tensor inversions compiled by
the CMT group (Ekstr\"om {\it et al.}, 2005).
The present catalog contains more than 28,000 earthquake
entries for the period 1977/1/1 to 2007/12/31.
Earthquake size is characterized by a scalar seismic moment
$M$.
The moment magnitude can be calculated from the seismic moment
(measured in Nm) value as
\be
m_W \ = \ (2/3) \cdot \log_{10} M - 6
\, .
\label{NBD_Eq36}
\ee
The magnitude threshold for the catalog is $m5.8$ (Kagan,
2003).

The PDE worldwide catalog is published by the USGS (U.S.\
Geological Survey, 2008); in its final form, the catalog
available at the time this article was written ended on
January 1, 2008.
The catalog measures earthquake size, using several magnitude
scales, and provides the body-wave ($m_b$) and surface-wave
($M_S$) magnitudes for most moderate and large events since
1965 and 1968, respectively.
The moment magnitude ($m_W$) estimate has been added recently.

Determining one measure of earthquake size for the PDE
catalog entails a certain difficulty.
For example, Kagan (1991) calculates a weighted average of
several magnitudes to use in the likelihood search.
Kagan (2003) also analyzes systematic and random errors for
various magnitudes in the PDE catalog.
At various times different magnitudes have been listed in the
PDE catalog, and establishing their relationships is
challenging.
Therefore, we chose a palliative solution: for each earthquake
we use the maximum magnitude among those magnitudes shown.
This solution is easier to carry out and the results are
easily reproducible.
For moderate earthquakes usually $m_b$ or $M_S$ magnitude is
selected.
For larger recent earthquakes the maximum magnitude
is most likely $m_W$.
Depending on the time period and the region, the magnitude
threshold of the PDE catalog is of the order 4.5 to 4.7
(Kagan, 2003).

The CalTech (CIT) dataset (Hileman {\it et al.}, 1973; Hutton
and Jones, 1993; Hutton {\it et al.}, 2006) was the first
instrumental local catalog to include small earthquakes ($m
\ge 3$), beginning with 1932/01/01.
The available catalog ends at 2001/12/31.
The magnitude threshold of the 1932-2001 catalog is close to
$m3$ [however, see Kagan (2004) for the threshold variations
after strong earthquakes].
In recent years, even smaller earthquakes have been included
in the catalog, so for the 1989-2001 time period, a threshold
of $m2$ is assumed.
We selected earthquakes in a spherical rectangular window
(latitudes $ > 32.5^\circ$N and $ \le 36.5^\circ$N,
longitudes $ > 114.0^\circ$W and $ \le 122.0^\circ$W)
to analyze.

\section {Earthquake numbers distribution
}
\label{numb}

\subsection {Statistical analysis of earthquake catalogs
}
\label{appl}

Several problems arise when the theoretical considerations of
Section~\ref{theo} are applied to earthquake catalogs.
Due to the limited sensitivity of a seismographic network and
its sparse spatial coverage, catalogs are incomplete in
magnitude, time, and space (Kagan, 2003).
In particular, the magnitude threshold of completeness varies
in time, usually decreasing during the catalog time-span.
Moreover, after strong earthquakes due to mainshock coda waves
and interference by other stronger events, small aftershocks
are usually absent for a period of a few hours to a few days
(Kagan, 2004).
In the best local catalogs this {\sl `dead'} period can be
significantly reduced (Enescu {\it et al.}, 2009), suggesting
that the lack of aftershocks in the interval larger than a few
minutes is an artifact of catalog incompleteness in the early
stages of an aftershock sequence.

An additional problem in comparing theoretical calculations to
observations is identifying earthquake clusters.
Due to delays in dependent events temporal distribution
described by Omori's law, clusters usually overlap in time and
space.
Only in zones of low tectonic deformation can aftershock
clusters of large earthquakes be distinguished.
Sometimes this aftershock sequence decay takes centuries (Ebel
{\it et al.}, 2000).
In more active regions, many earthquake clusters overlap.

It is possible, in principle, to use stochastic declustering
in defining statistical interrelations between various events.
For the branching-in-magnitude (Fig.~\ref{fig01}a) such a
procedure was first applied by Kagan and Knopoff (1976,
see Table~XVIII); for the branching-in-time model, Kagan and
Jackson (1991) and Zhuang {\it et al.}\ (2004) proposed the
identification procedure.
However, such stochastic decomposition is typically
non-unique.
It depends on the details of a stochastic model
parametrization and has not been attempted in this work.
For branching-in-time models, connections of small vs large
events (such as foreshock-mainshock ties) create an additional
difficulty in stochastic reconstruction.
See more discussion on this topic at the end of
Subsection~\ref{gen1}.
Handling these connections unambiguously is difficult.

Moreover, the equations derived in Subsection~\ref{gen1}
neglect temporal and spatial parameters of earthquake
occurrence.
Hence, they are valid for large space-time windows
exceeding the size of the largest earthquake cluster.
The time-span of available earthquake catalogs is very
limited.
For the largest earthquakes, the duration of the aftershock
sequences is comparable or exceeds a typical catalog length.
Additionally, when studying earthquake number distribution,
we need to subdivide the catalog into several sub-units, thus
reducing again the temporal or spatial window.
Furthermore, the theoretical model neglects long-term
modulations of seismicity (Kagan and Jackson, 1991; Lombardi
and Marzocchi, 2007).
Therefore, we should not expect close agreement between the
theoretical formula and empirical results.
Only general regularities in distribution behavior can be
seen.

\subsection {Observed earthquake numbers distributions
}
\label{dist}

\begin{figure}
\begin{center}
\includegraphics[width=0.75\textwidth]{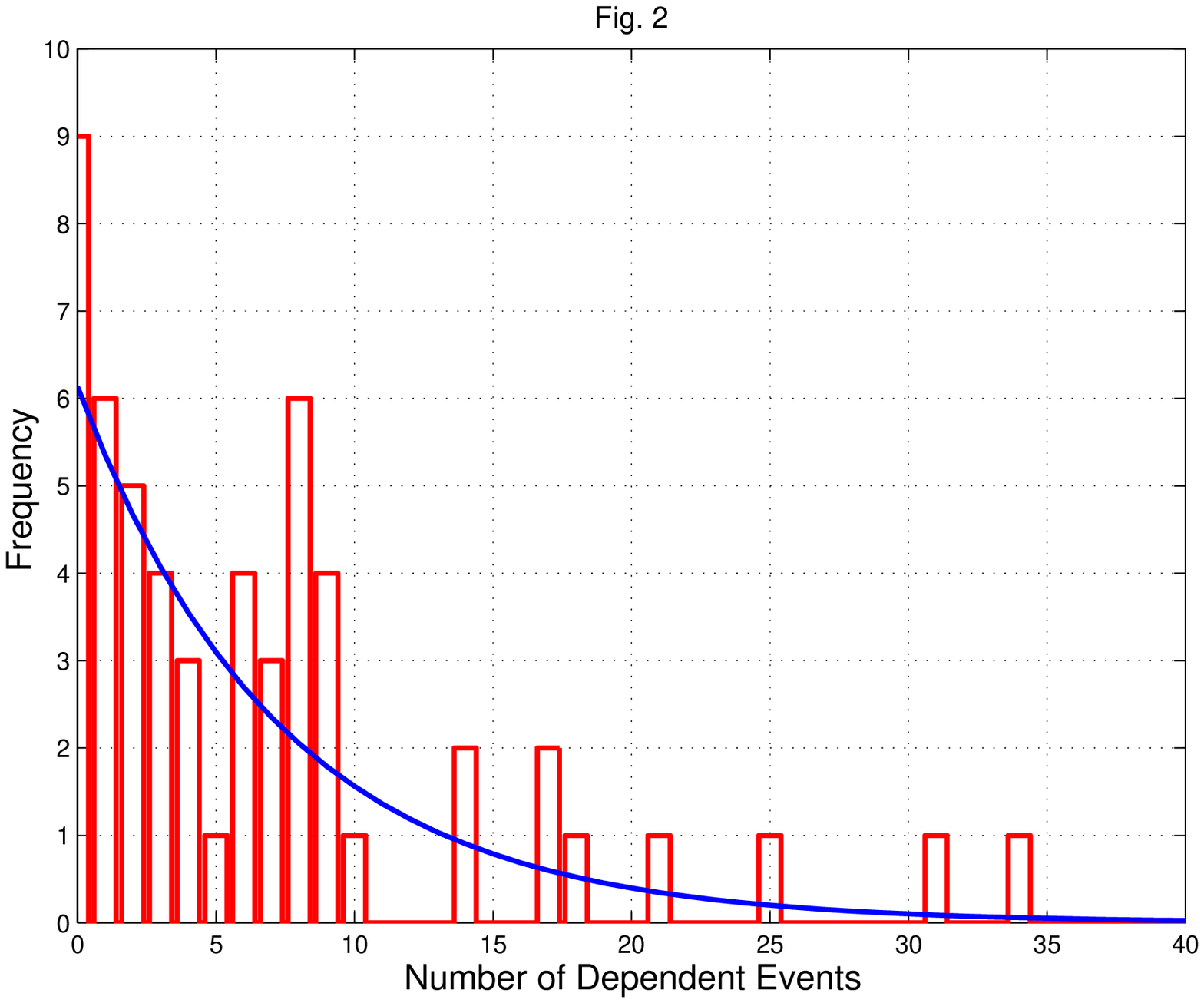}
\caption{\label{fig02}
}
\end{center}
Red bars -- distribution of aftershock numbers $m \ge 4.7$ in
the PDE catalog 1977-2007, following $7.2 > m_W \ge 7.1$
earthquakes in the CMT catalog.
The blue line is an approximation of the distribution using
the geometric law (Eqs.~\ref{NBD_Eq05} and \ref{NBD_Eq06})
with the parameter $\hat p = 0.1279$, calculated using
(\ref{NBD_Eq26}).
\end{figure}

\begin{figure}
\begin{center}
\includegraphics[width=0.75\textwidth]{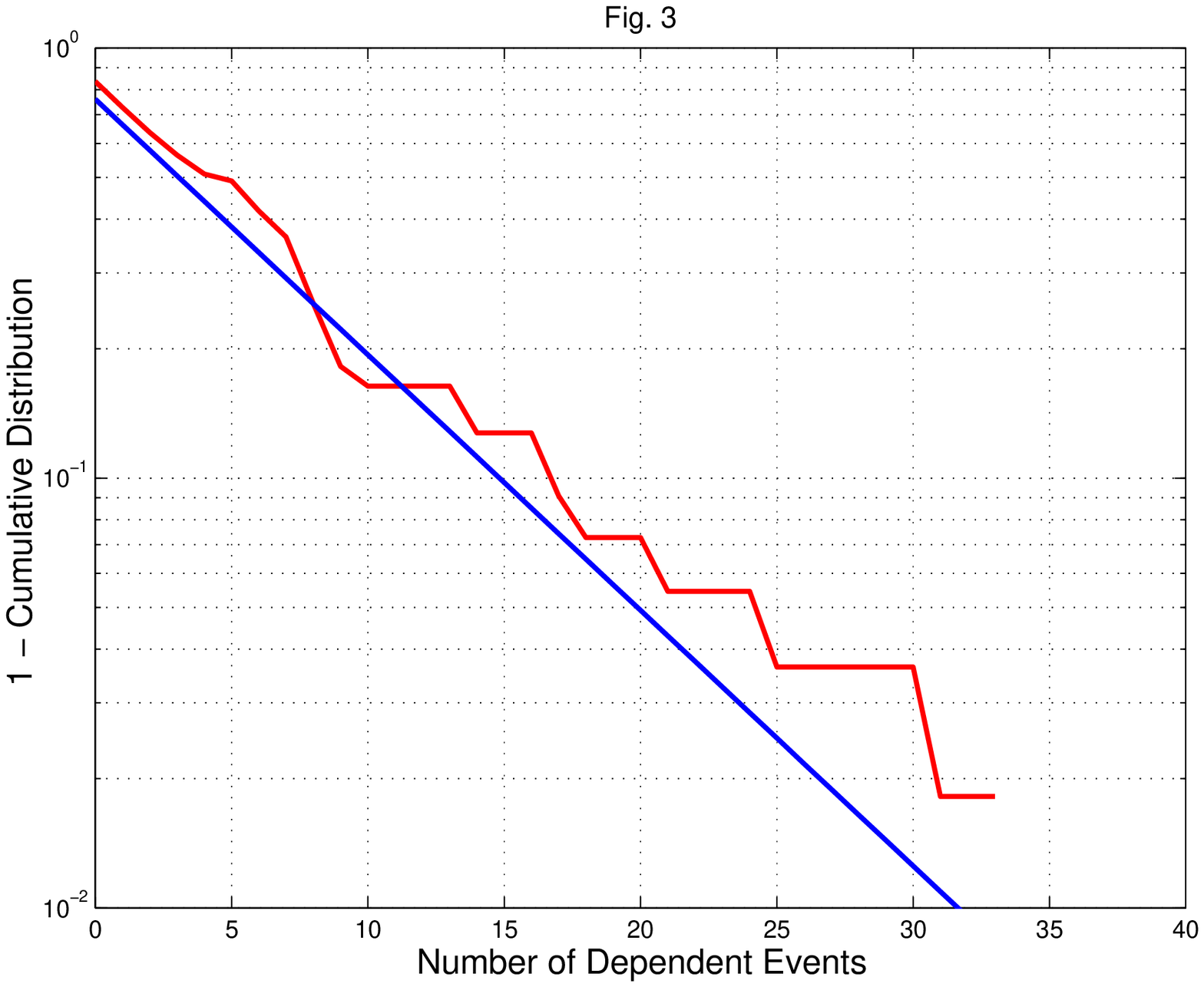}
\caption{\label{fig03}
}
\end{center}
Red line -- survival function (1 - Cumulative distribution)
of aftershock numbers $m
\ge 4.7$ in the PDE catalog 1977-2007, following $7.2 > m_W
\ge 7.1$ earthquakes in the CMT catalog.
The blue line is an approximation of the distribution using
the geometric law (Eqs.~\ref{NBD_Eq05} and \ref{NBD_Eq06}).
\end{figure}

Fig.~\ref{fig02} shows the distribution of shallow (depth
0-70~km) aftershock numbers in the PDE catalog, following
a $m7.1-7.2$ event in the CMT catalog.
We count the aftershock number during the first day within a
circle of radius $R$ (Kagan, 2002)
\be
R (m) \ = \ 20 \cdot 10^{m-6} \, \quad {\rm km}
\, .
\label{NBD_Eq37}
\ee
Even taking into account location errors in both catalogs, the
radius of 200~km for the $m7$ event guarantees that, almost
all the first day aftershocks will be counted.
The geometric distribution curve seems to approximate the
histogram satisfactorily.
Comparing the observed cumulative distribution with its
approximation in Fig.~\ref{fig03} also confirms that the
geometric law appropriately describes the aftershock number
distribution.

\begin{figure}
\begin{center}
\includegraphics[width=0.75\textwidth]{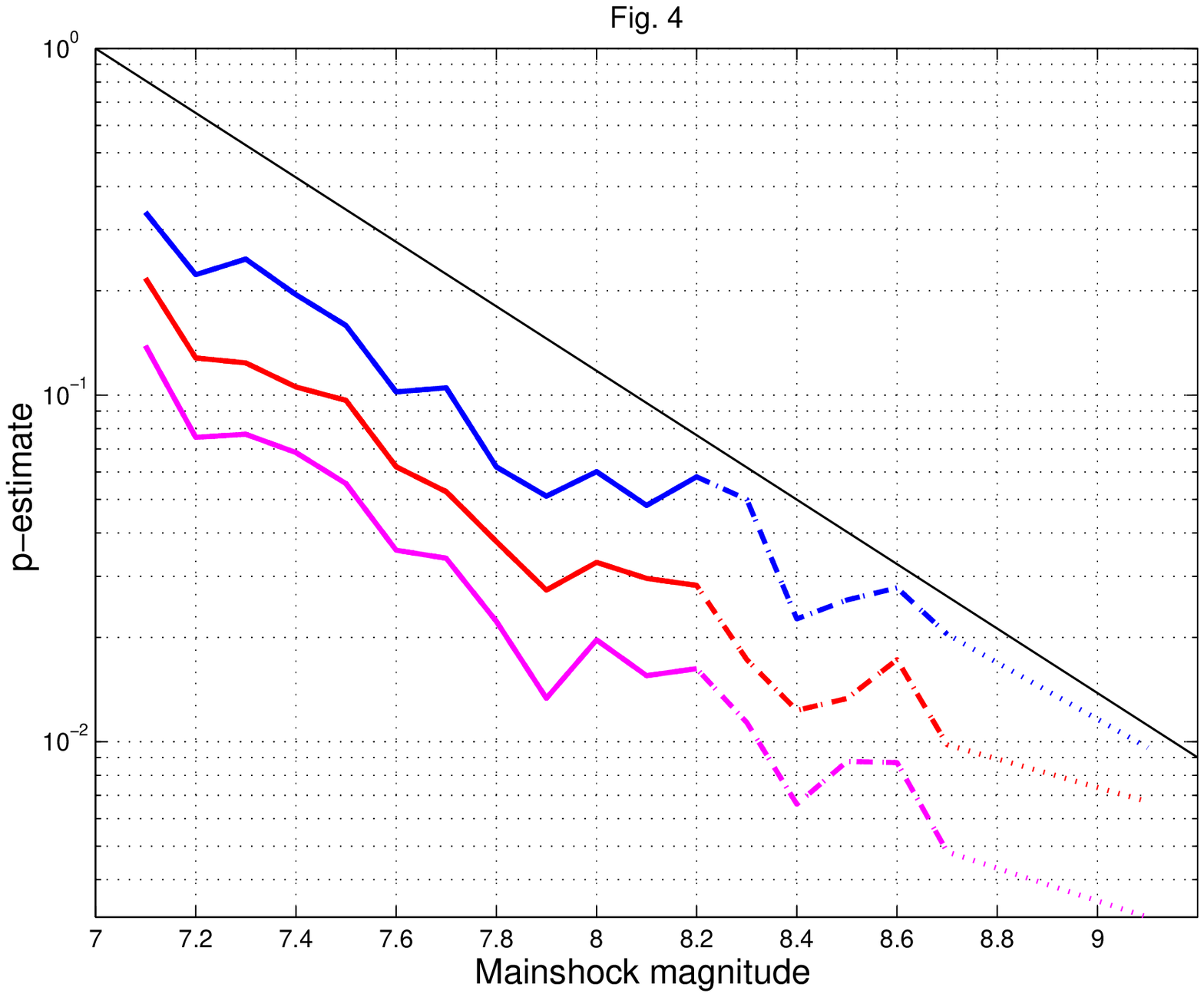}
\caption{\label{fig04}
}
\end{center}
Dependence of the $\hat p$-estimate on mainshock magnitude.
The blue line is for $m5.0$ aftershocks during 1-day interval
after a mainshock,
the red line is for $m4.7$ aftershocks during 1-day interval
after a mainshock, and
the magenta line is for $m4.7$ aftershocks during a 7-day
interval after a mainshock.
Solid lines connect data points with more than 3 mainshocks,
dashed lines are for a single mainshock, and dotted lines
connect the estimate for 2004 $m9.1$ Sumatra earthquake.
The thin black line corresponds to $p \propto 10^{- 1.5 \,
m \, \beta}$ (see Eq.~\ref{NBD_Eq04}).
\end{figure}

For the geometric distribution, Fig.~\ref{fig04} shows the
dependence in the $p$-parameter on the mainshock magnitude.
The $\hat p$-values decay approximately by a factor of $10^{
\, - \, 1.5 \, \beta} $ with a magnitude increase by one
unit.
This behavior is predicted by Eqs.~\ref{NBD_Eq04} and
\ref{NBD_Eq07}.

\begin{figure}
\begin{center}
\includegraphics[width=0.75\textwidth]{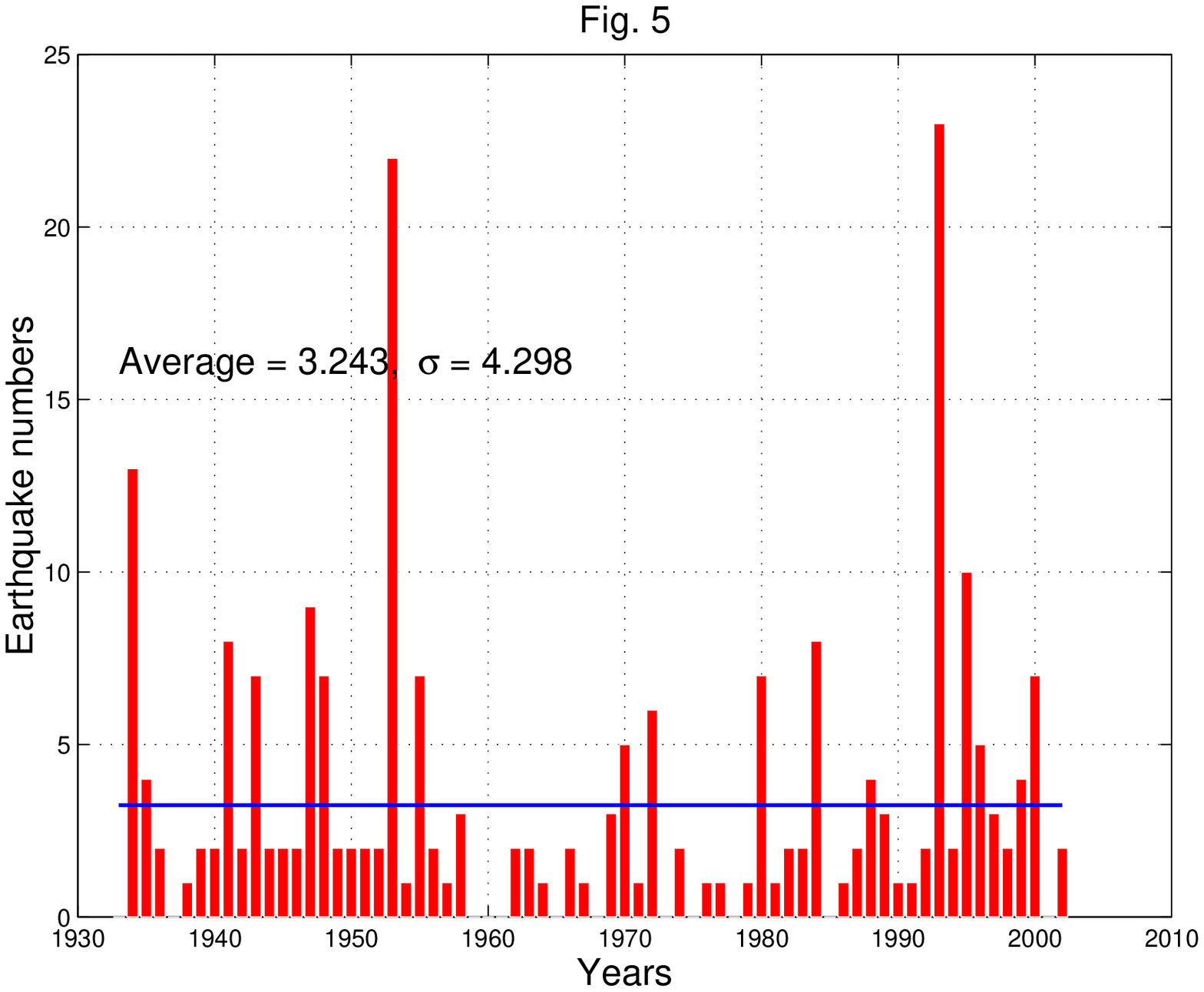}
\caption{\label{fig05}
}
\end{center}
Annual numbers of earthquakes $m \ge 5.0$ in southern
California, 1932-2001.
\end{figure}

Fig.~\ref{fig05} displays an example of earthquake numbers in
equal time intervals (annual in this case) for the CIT
catalog (see also Fig.~5 by Kagan and Jackson, 2000).
Even a casual inspection suggests that the numbers are
over-dispersed compared to the Poisson process: the standard
deviation is larger than the average.
The number peaks are easily associated with large earthquakes
in southern California:
the $m7.5$ 1952 Kern County event,
the $m7.3$ 1992 Landers, and
the $m7.1$ 1999 Hector Mine earthquakes.
Other peaks can usually be traced back to strong
mainshocks with extensive aftershock sequences.

\begin{figure}
\begin{center}
\includegraphics[width=0.75\textwidth]{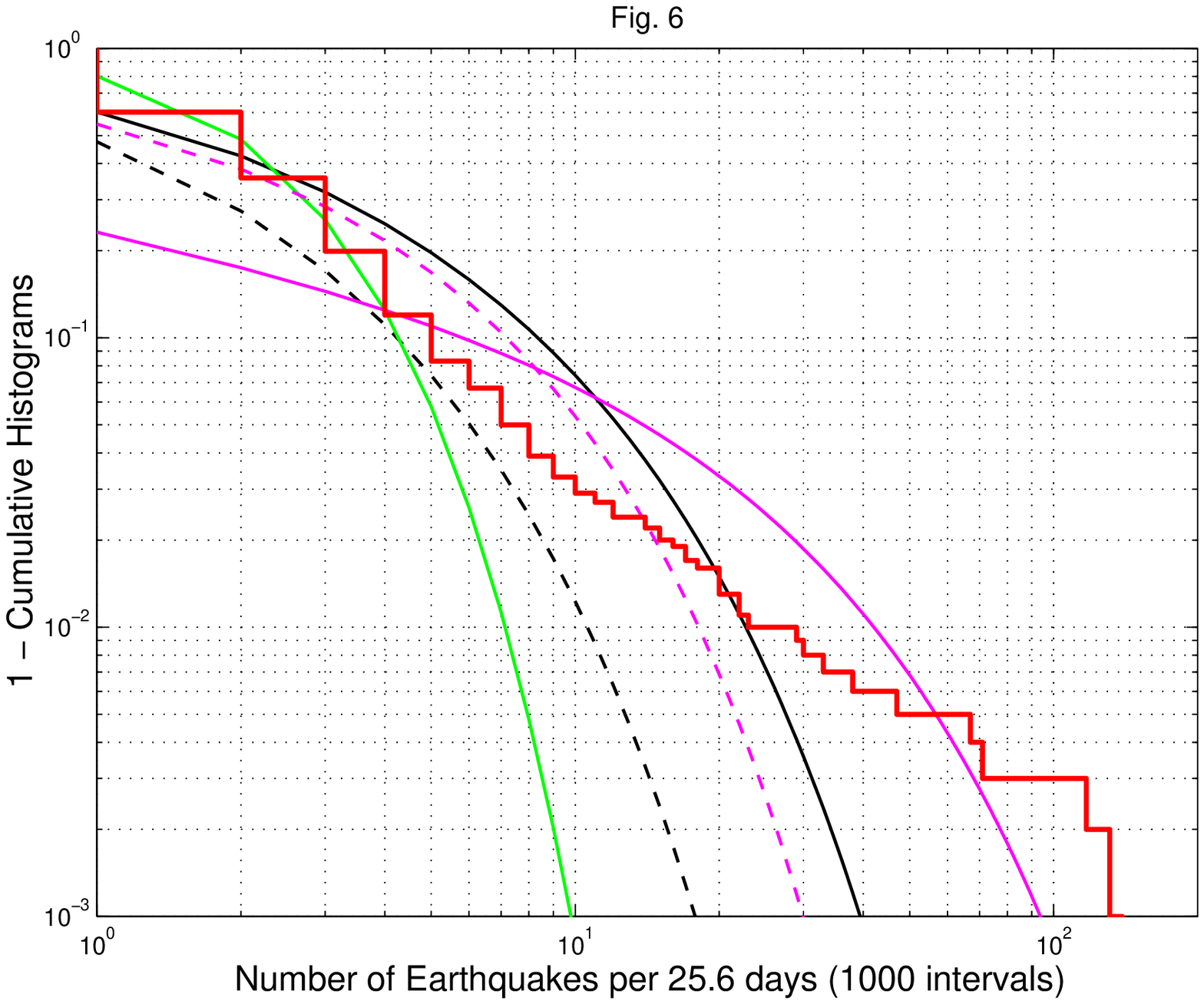}
\caption{\label{fig06}
}
\end{center}
Survival function (1 - Cumulative distribution) of earthquake
numbers for the CIT catalog 1932-2001, $m~\ge~4.0$.
The step-function shows the observed distribution in 25.6 days
time intervals (1000 intervals for the whole time period).
The green curve is the approximation by the Poisson
distribution (\ref{NBD_Eq14}, \ref{NBD_Eq38});
the magenta solid line is the NBD approximation
(Eqs.~\ref{NBD_Eq15}, \ref{NBD_Eq18}, and \ref{NBD_Eq39}),
with parameters estimated by the MLE method; the magenta
dashed curve is the NBD approximation with parameters are
estimated by the moment method (see Subsection~\ref{stat});
the black dashed line is the approximation by the logarithmic
distribution (\ref{NBD_Eq10});
the black solid line is the approximation by the logarithmic
law for the zero-truncated distribution.
\end{figure}

In large time intervals, one would expect a mix of several
clusters, and according to Eq.~\ref{NBD_Eq09} the
numbers would be distributed as the NBD.
At the lower tail of the number distribution the small time
intervals may still have several clusters due to weaker
mainshocks.
However, one small time interval would likely contain only one
large cluster.
Therefore, their distribution would be approximated by the
logarithmic law (\ref{NBD_Eq10}-\ref{NBD_Eq11}).
Fig.~\ref{fig06} confirms these considerations.
The upper tail of the distribution resembles a power-law with
the exponent close to 1.0.

The observed distribution in Fig.~\ref{fig06} is compared
to several theoretical curves ({\it cf.}\ Kagan, 1996).
The Poisson cumulative distribution is calculated using
the following formula
\be
F(k) \ = \ P(N < k) \ = \ {1 \over {k!}} \,
\int\limits_{\lambda}^{\infty} \, y^k \, e^{-y} \, dy \ = \ 1
\, - \, \Gamma (k+1, \, \lambda),
\label{NBD_Eq38}
\ee
where $\Gamma (k+1, \, \lambda)$ is an incomplete gamma
function.
For the NBD
\be
F(k) \ = \ P(N < k) \ = \ {1 \over {B(\tau, \, k+1)}} \,
\int\limits_{0}^{\theta} \, y^{\tau - 1} \, (1 - y)^k \, dy,
\label{NBD_Eq39}
\ee
where $B(\tau, \, k+1)$ is a beta function.
The right-hand part of the equation corresponds to an
incomplete beta function, $B (\tau, \, k+1, \, x)$ (Gradshteyn
and Ryzhik, 1980).

For the logarithmic distribution (\ref{NBD_Eq10}) two
parameter evaluations are made: one based on the naive average
number counts and the other on number counts for a
`zero-truncated' empirical distribution (Patil, 1962,
p.~70; Patil and Wani, 1965, p.~386).
The truncation is made because the logarithmic distribution
is not defined for a zero number of events.
Thus, we calculate the average event count for only 60\%
of intervals having a non-zero number of earthquakes.

These theoretical approximations produce an inadequate fit to
the observation.
The Poisson law fails because there are strong clusters in the
catalog.
The NBD fails for two reasons:
the clusters are truncated in time and the cluster mix is
insufficient, especially at the higher end of the
distribution.
Moreover, as we mentioned above, the long-term seismicity
variations may explain the poor fit.
The logarithmic distribution fails at the lower end, since
several clusters, not a single one, as expected by the
distribution, are frequently present
in an interval.
The quasi-exponential tail (see Eq.~\ref{NBD_Eq12}) is not
observed in the plot, since in the CIT catalog there are no
events with a magnitude approaching the maximum (corner)
magnitude.
For California the corner magnitude should be on the order of
$m8$ (Bird and Kagan, 2004).

We produced similar plots for other global catalogs (the PDE
and CMT); and the diagrams also display a power-law tail for
small time intervals.
However, this part of a distribution is usually smaller.
This finding is likely connected to the smaller magnitude
range of these catalogs; fewer large clusters of dependent
events are present in these datasets.

\begin{figure}
\begin{center}
\includegraphics[width=0.75\textwidth]{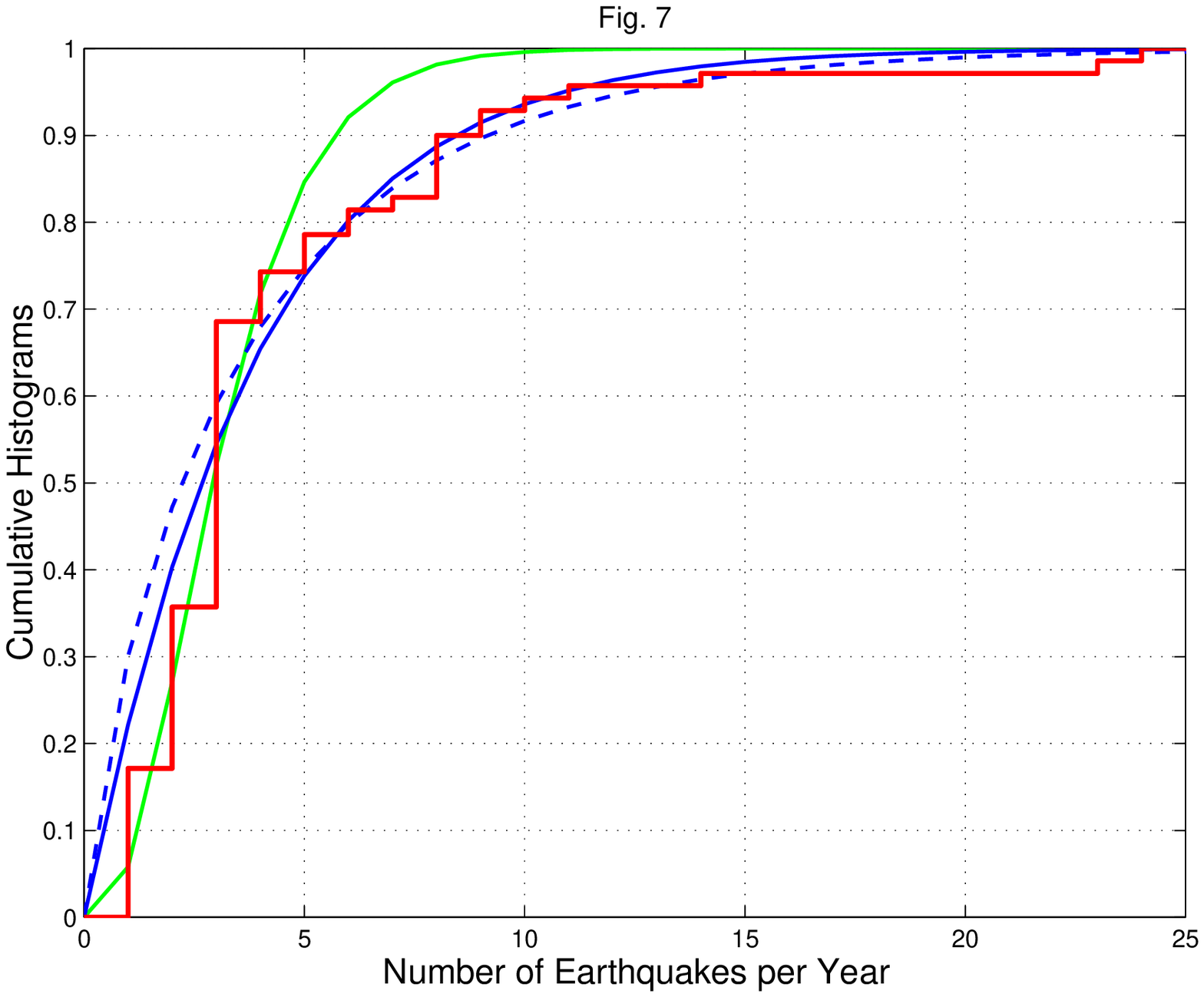}
\caption{\label{fig07}
}
\end{center}
Cumulative distribution of annual earthquake numbers
for the CIT catalog 1932-2001, $m~\ge~5.0$.
The step-function shows the observed distribution, and
the solid green curve is the theoretical Poisson distribution
(\ref{NBD_Eq38}).
Two negative binomial curves (\ref{NBD_Eq39}) are also
displayed: for the dashed curve the parameters $\theta$ and
$\tau$ are evaluated by the moment method, for the solid
curve MLEs are used.
The negative binomial curves fit the tails much better than
the Poisson does.
\end{figure}

Fig.~\ref{fig07} shows a cumulative distribution for annual
earthquake numbers.
Again the fit by the Poisson law (\ref{NBD_Eq38}) is poor,
whereas the NBD (\ref{NBD_Eq39}) is clearly the better
approximation.

\subsection {Likelihood analysis
}
\label{like}

\begin{figure}
\begin{center}
\includegraphics[width=0.75\textwidth]{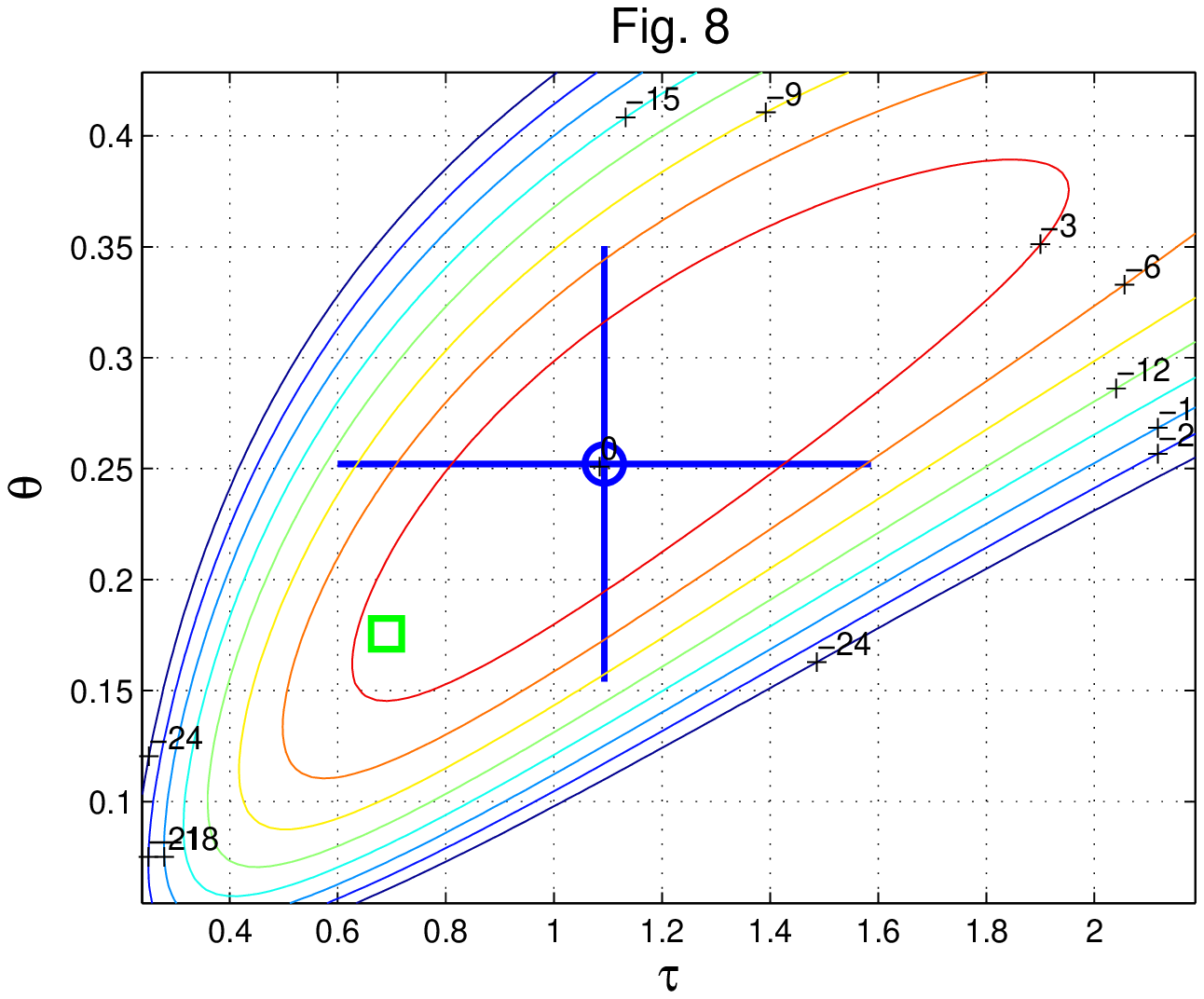}
\caption{\label{fig08}
}
\end{center}
The log-likelihood function map for the
CIT earthquake catalog 1932-2001, $m \ge 5.0$, annual event
numbers are analyzed.
The standard representation of the NBD (\ref{NBD_Eq15}) is
used.
The green square is the estimate of $\theta$ and $\tau$
by the moment method, whereas blue circle shows the MLE
parameter values.
An approximate $95\%$-confidence area, based on asymptotic
relations, corresponds to the contour labeled $-3.0$.
Two orthogonal line intervals centered at the circle are 95\%
confidence limits for both parameters, obtained by {\scaps
matlab} (Statistics Toolbox).
The correlation coefficient $\rho$ between these estimates
(also evaluated by {\scaps matlab}) is 0.867.
\end{figure}

Fig.~\ref{fig08} displays a two-dimensional plot of the
log-likelihood function (\ref{NBD_Eq35}) for the standard
version of the NBD.
Such plots work well for parameter estimates: if the relation
is non-linear or the parameter value needs to be restricted
(if, for example, it goes into the negative domain or to
infinity, etc.) such plots provide more accurate information
than does the second-derivative matrix.

The diagram shows that the parameter estimates are highly
correlated.
Moreover, isolines are not elliptical, as required by the
usual asymptotic assumption; thus, uncertainty estimates based
on the Hessian matrix (see Eq.~\ref{NBD_Eq35}) may be
misleading.
The 95\% confidence limits obtained by the {\scaps matlab}
(Statistics Toolbox) procedure testify to this.
Wilks (1962, Chap.\ 13.8) shows that the log-likelihood
difference is asymptotically distributed as $ {1 \over 2 }
\chi^2(2)$ (chi-square distribution with two degrees of
freedom, corresponding to two parameters of the NBD model).
Thus, the isoline [$-3.0$] at the log-likelihood map should
approximately equal 95\% confidence limits.
The moment estimates (Eqs.~\ref{NBD_Eq27} and \ref{NBD_Eq28})
are within the 95\% limits.

\begin{figure}
\begin{center}
\includegraphics[width=0.75\textwidth]{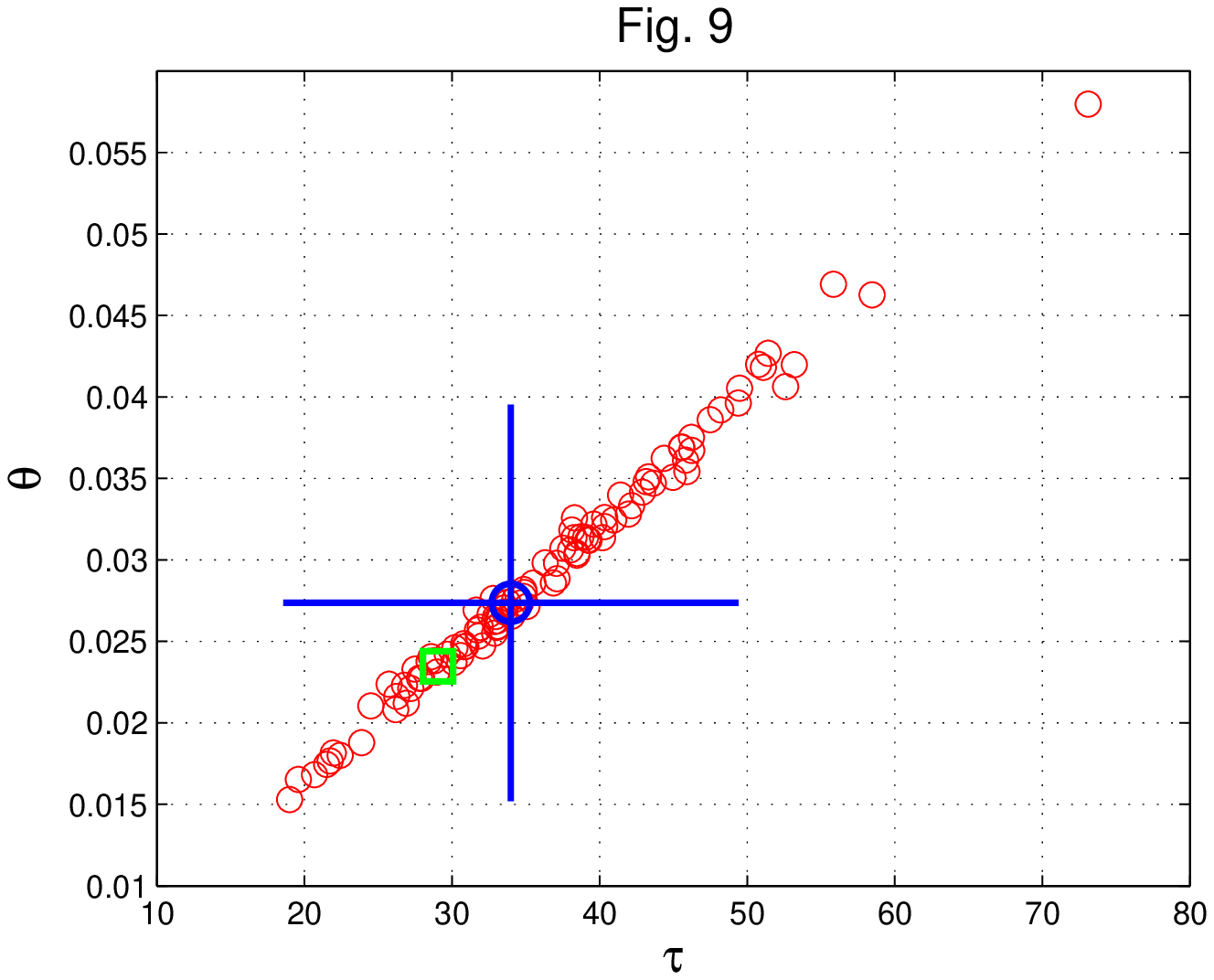}
\caption{\label{fig09}
}
\end{center}
The standard NBD parameter calculations for
PDE earthquake catalog 1969-2007, $m \ge 5.0$, annual event
numbers are analyzed.
The large green square is the estimate of $\theta$ and $\tau$
by the moment method, whereas large blue circle shows the MLE
parameter values:
$\tau = 33.98 \pm 15.42$ and $\theta = 0.0274 \pm 0.0122$.
Two orthogonal line intervals centered at the circle are 95\%
confidence limits for both parameters.
Small circles are simulated parameter estimates, using MLEs.
In simulations the parameter estimates for $\theta$ and $\tau$
are also MLEs (see above).
\end{figure}

For the PDE catalog, where $\rho = 0.993$, the effect of the
high correlation coefficient ($\rho$) on parameter estimates
of the standard NBD is demonstrated more strongly in
Fig.~\ref{fig09}.
In addition to the parameters for the empirical frequency
numbers, estimated by the MLE and the moment method, parameter
estimates for 100 simulations produced, applying the {\scaps
matlab} package, are also shown here.
These simulations used the MLEs as their input.
The parameters scatter widely over the plots, showing that
regular uncertainty estimates cannot fully describe their
statistical errors.

Similar simulations with other catalogs and other magnitude
thresholds show that for the standard NBD representation,
simulation estimates are widely distributed over $\tau, \,
\theta$ plane.
Depending on the original estimation method, the moment or the
MLE, the estimates are concentrated around parameter values
used as the simulation input.

\begin{figure}
\begin{center}
\includegraphics[width=0.75\textwidth]{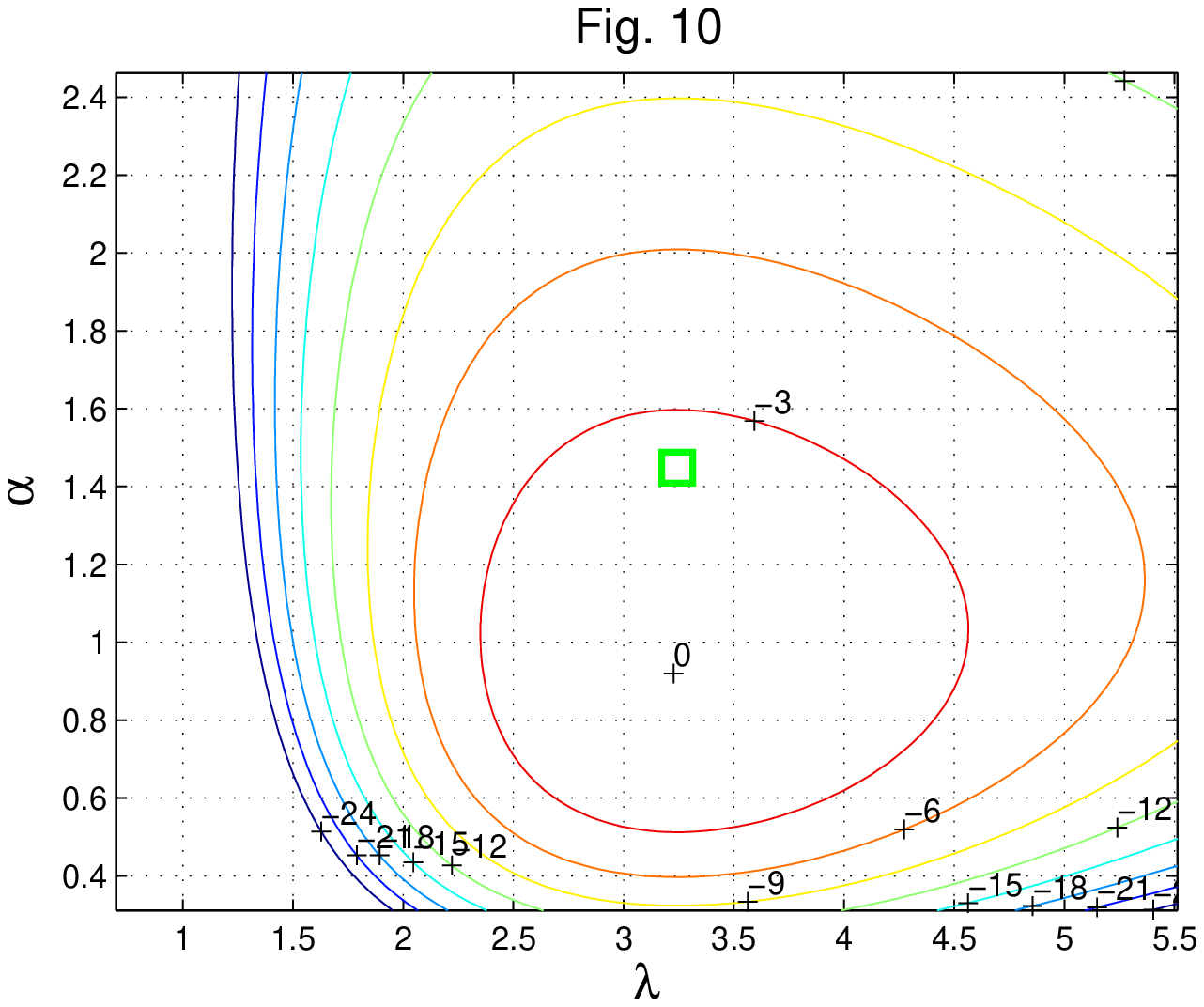}
\caption{\label{fig10}
}
\end{center}
The log-likelihood function map for
CIT earthquake catalog 1932-2001, $m \ge 5.0$, annual event
numbers are analyzed.
The alternative representation of the NBD (\ref{NBD_Eq18}) is
used.
The green square is the estimate of $\theta$ and $\lambda$ by
the moment method.
An approximate $95\%$-confidence interval, based on asymptotic
relations, corresponds to the contour labeled $-3.0$.
\end{figure}

\begin{figure}
\begin{center}
\includegraphics[width=0.75\textwidth]{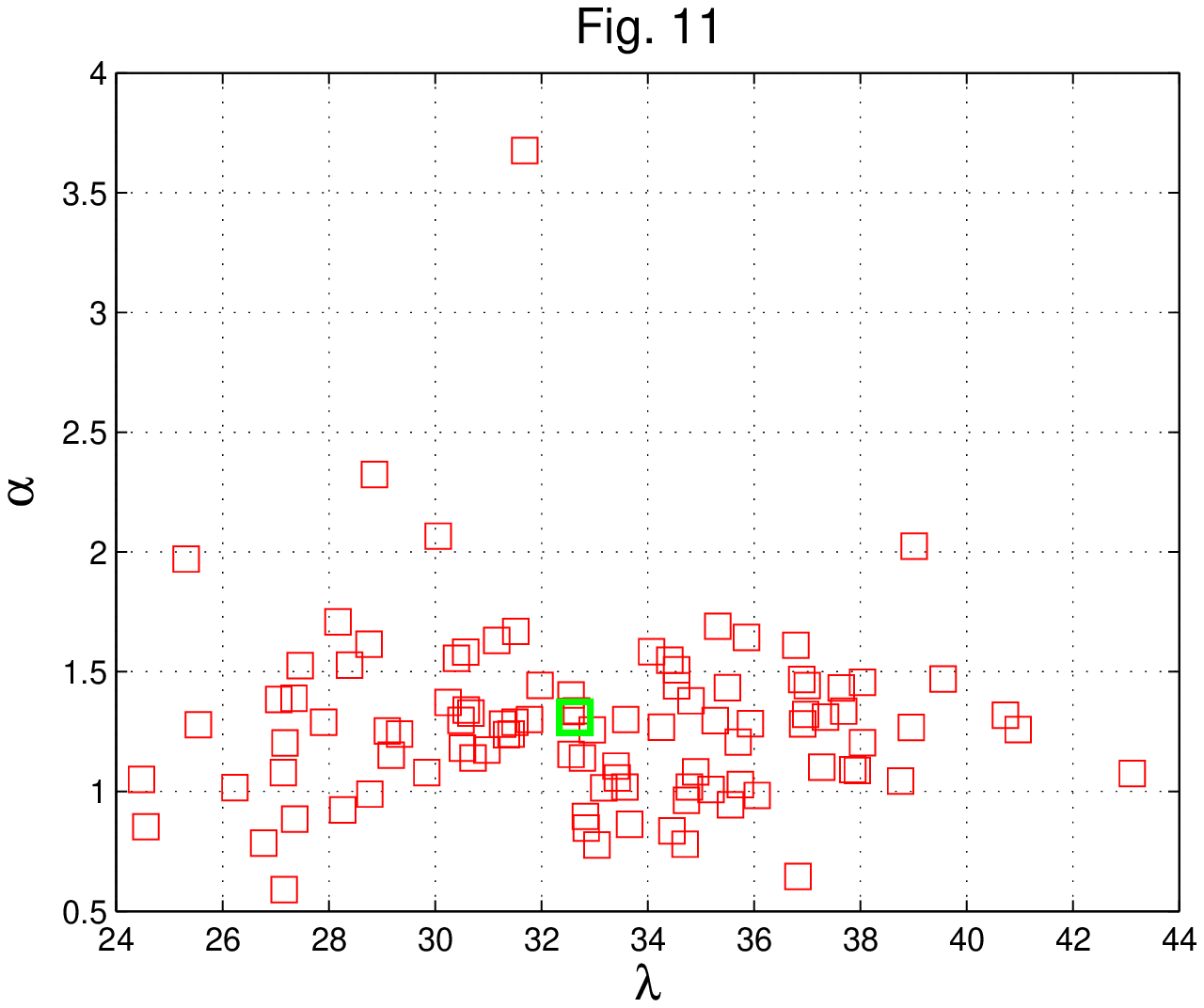}
\caption{\label{fig11}
}
\end{center}
Simulation results for Caltech earthquake catalog 1932-2001,
$m \ge 4.0$, annual event numbers are analyzed.
The alternative representation of the NBD (\ref{NBD_Eq18}) is
used.
The large green square is the estimate of $\alpha $ and
$\lambda$ by the moment method for the catalog data.
Small squares are simulated parameter estimates, using
$\alpha $ and $\lambda$ as input.
In these displays the parameters $\alpha $ and $\lambda$
are moment estimates.
\end{figure}

Fig.~\ref{fig10} displays the likelihood map for the
alternative form (\ref{NBD_Eq18}) of the NBD.
It shows a different behavior from that of Fig.~\ref{fig08}.
Isolines are not inclined in this case, indicating that the
correlation between the parameter estimates should be slight.
The simulation results shown in Fig.~\ref{fig11} confirm this.

\begin{figure}
\begin{center}
\includegraphics[width=0.75\textwidth]{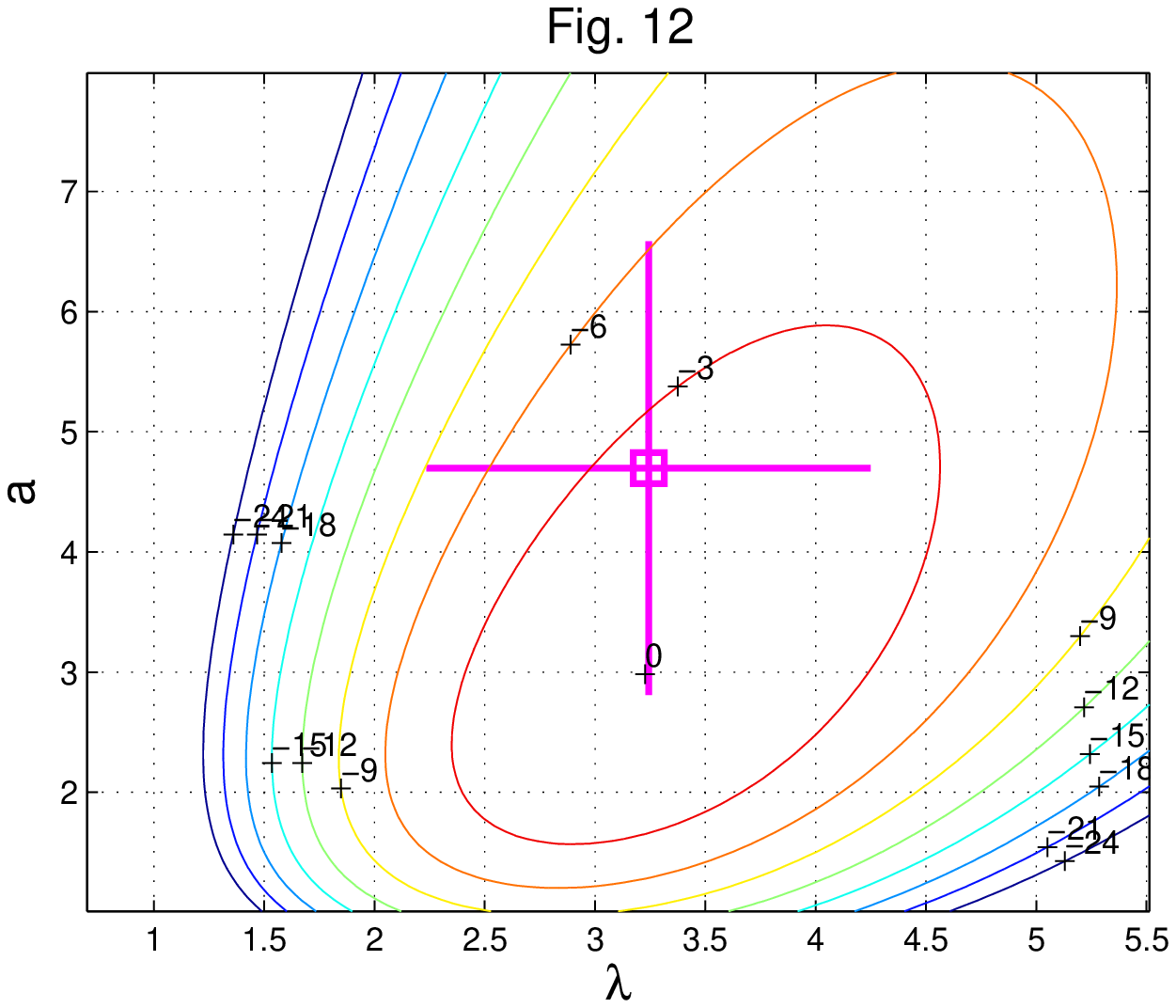}
\caption{\label{fig12}
}
\end{center}
The likelihood function map for
CIT earthquake catalog 1932-2001, $m \ge 5.0$, annual event
numbers are analyzed.
Evans' (1953) representation of the NBD (\ref{NBD_Eq20})
is used.
The red square is the moment estimate of $a$ and
$\lambda $.
An approximate $95\%$-confidence interval, based on asymptotic
relations, corresponds to the contour labeled $-3.0$.
Two orthogonal line intervals centered at the square are 95\%
confidence limits for both parameters, based on Evans' (1953)
variance formula for moment estimates.
\end{figure}

Fig.~\ref{fig12} shows the likelihood map for Evans'
distribution.
Again the isolines are less inclined with regard to axes,
showing a relatively low correlation between the estimates.
Using formulas (\ref{NBD_Eq32}--\ref{NBD_Eq34}), supplied by
Evans (1953, p.~203) we calculate 95\% uncertainties shown in
the plot.
The correlation coefficient ($\rho$) between the estimates is
$\sim 0.15$.

Fig.~\ref{fig13} shows the dependence of the log-likelihood
difference $\ell - \ell_0$ on the magnitude threshold
($\ell_0$ is the log-likelihood for the Poisson distribution,
see Eq.~\ref{NBD_Eq35}).
The difference increases as the threshold decreases,
testifying again that large earthquakes are more Poisson.
The Poisson distribution is a special case of the NBD.
Therefore, we can estimate at what log-likelihood difference
level we should reject the former hypothesis as a model of
earthquake occurrence.
Wilks (1962; see also Hilbe, 2007) shows that the
log-likelihood difference is distributed for a large number of
events as $ {1 \over 2 } \chi^2(1)$ (chi-square distribution
with one degree of freedom.
This corresponds to one parameter difference between the
Poisson and NBD models).
The 95\% confidence level corresponds to the $\ell - \ell_0$
value of 1.92.
Extrapolating the curve suggests that earthquakes smaller than
$m6.5$ in southern California cannot be approximated by a
Poisson model.
If larger earthquakes were present in a catalog, these events
($m \ge 6.5$) might be as clustered, so this threshold would
need to be set even higher.

\subsection {Tables of parameters
}
\label{tabl}

Tables~\ref{Table1}--\ref{Table3} show brief results of
statistical analysis for three earthquake catalogs.
These tables display three dependencies of NBD parameter
estimates: (a) on the magnitude threshold
($m_c$); (b) on time intervals ($\Delta T$) a catalog
time-span is subdivided; and (c) on a subdivision of a catalog
space window.
Three NBD representations are investigated: the standard, the
alternative, and Evans' formula.
Since the parametrization of the last two representations is
similar, we discuss below only the standard and the
alternative set of parameters.
We used the moment method, which is more convenient in
applications, to determine the parameter values.

The parameter variations in all subdivisions exhibit
similar features:
\hfil\break
$\bullet$
(a)
In the PDE catalog the parameter $\alpha$ decreases as the
threshold, $m_c$, increases ($\alpha \to 0$ corresponds to the
Poisson distribution).
The $\theta$-value displays the opposite behavior:
when $\theta \to 1.0$, the NBD approaches the Poisson law.
The $\theta$ parameter shows a similar behavior in the CMT
and CIT catalogs (the negative parameter values for $m_c =
7.0$ in the CMT catalog reveal that the NBD lacks appropriate
fit for our observations).
The $\alpha$ parameter displays no characteristic
feature for the CMT and CIT catalogs.
The $m_c = 2.0$ results for the CIT catalog are obtained for
the period 1989-2001, so they cannot be readily compared to
results for other magnitude thresholds.

Fig.~\ref{fig14} displays for all catalogs the dependence of
$\theta$ on the magnitude.
The decay of the parameter values can be approximated by
a power-law function; such behavior can be explained by
comparing Eqs.~(\ref{NBD_Eq09}) and (\ref{NBD_Eq21}).
This diagram is similar to Fig.~6 in Kagan and Jackson (2000),
where the parameter $\theta$ is called $\Upsilon$, and the
magnitude/moment transition is as shown in (\ref{NBD_Eq36}).
\hfil\break
$\bullet$
(b)
The $\theta$-values are relatively stable for both
global and the CIT catalogs, showing that the event number
distribution does not change drastically as the time interval
varies.
It is not clear why the $\alpha$ significantly changes as the
time intervals decrease.

When the $\Delta T$ changes, the behaviour of the $\alpha$ and
the $\theta$, is generally contradictory: both parameters
increase with the decrease of the time interval.
This trend is contrary to the dependence of these variables on
the magnitude threshold [see item (a) above].
The $\theta$ increase may suggest that the distribution
approaches the Poisson law as $\Delta T \ \to \ 0$, whereas
the $\alpha$ trend implies an increase in clustering.
Such an anomalous behaviour is likely to be caused by a change
of the earthquake number distribution for small time
intervals.
Fig.~\ref{fig06} demonstrates that the upper tail of the
distribution is controlled by the logarithmic series law
(\ref{NBD_Eq10}).
Although the logarithmic distribution is a special case of the
NBD, it requires one parameter {\it versus} NBD's two degrees
of freedom.
Therefore, it is possible that when the logarithmic
distribution is approximated by the NBD, the parameters
$\alpha$ and $\theta$ of two different NBD representations
behave differently.
The dependence of the distribution parameters on the time
sampling interval needs to be studied from both theoretical
and observational points of view.

Fig.~\ref{fig15} displays the $\theta$ parameter behavior for
changing time intervals.
There is no regular pattern in the curves properties for the
three catalogs.
However, the $\theta$-values for the 1-year and 5-year
intervals can be used in testing earthquake forecasts for
these catalogs.
\hfil\break
$\bullet$
(c)
Two types of spatial subdivision are shown in the tables.
For global catalogs we subdivide seismicity into five
zones according to the tectonic deformation which prevails in
each zone (Kagan {\it et al.}, 2009).
The CIT catalog was subdivided into four geographic zones.
We also made a similar subdivision for the global catalogs
(not shown).
In all these subdivisions both $\alpha$ and $\theta$ are
relatively stable, whereas a sum of parameter
$\tau$-values for the subdivided areas approximately equals
the parameter's value for the whole area.
For example, the $\tau$-values for the CIT catalog are:
$1.0888 \, \approx \, 0.5566+0.235+0.265+0.5672 \ (1.6238)$.
By definition, the $\lambda$ parameter equals the sum for the
sub-areas.

\section {Discussion
}
\label{disc}

We studied theoretical distributions of earthquake counts.
The obtained discrete distributions (Poisson, geometric,
logarithmic, and NBD) are applied to approximate the event
numbers in earthquake catalogs.
The Poisson distribution is appropriate for earthquake numbers
when the catalog magnitude range (the difference between the
maximum and the threshold magnitudes) is small.
The geometric law applies for the earthquake number
distribution in clusters (sequences) with a fixed magnitude
range, whereas the logarithmic law is valid for all earthquake
clusters.
Finally, the NBD approximates earthquake numbers for extensive
time-space windows if the magnitude range is relatively large.

Our results can be used in testing earthquake forecasts to
infer the expected number of events and their confidence
limits.
The NBD parameter values shown in
Tables~\ref{Table1}--\ref{Table3} could be so used.
The 95\% uncertainties can be calculated by using
Eqs.~(\ref{NBD_Eq38}) and (\ref{NBD_Eq39}).

We have shown that the different implementations of the NBD
have dissimilar characteristics in the likelihood space of the
parameter estimates (Figs.~\ref{fig08}-\ref{fig12}).
The alternative (\ref{NBD_Eq18}) or the Evans (\ref{NBD_Eq20})
distributions clearly have parameter estimates that look as
non-correlated, and on that basis they are preferable
in the practical use, though additional investigations of
their performance in real earthquake catalogs should be
performed.

The moment method based estimates discussed in the paper
fall into a high confidence level, thus these estimates are
relatively effective.
They are much simpler to implement and faster in terms of
computational speed.
However, their uncertainties need to be determined.
Apparently such an evaluation can be done without much
difficulty.

As we mentioned in the Introduction, many studies have been
published on certain aspects of earthquake number
distributions.
But this work is first in combining the theoretical derivation
with statistical analysis of major features of earthquake
occurrence.
As a pioneer paper, it cannot resolve several issues related
to earthquake number distributions.
We list some of them below, hoping that other researchers,
both geophysicists and statisticians, will find their
solutions.

$\bullet$
1) Many dependent events are not included in earthquake
catalogs: a close foreshock may be misidentified as an initial
stage of larger shock, rather than an individual event.
The coda waves of a large event and strong aftershocks hinder
identification of weaker earthquakes.
It is likely that these phenomena can be modeled by a more
sophisticated scheme of population processes (such as
age-dependent branching processes, etc.) and a more
detailed quantitative derivation can be obtained.

$\bullet$
2) It would be interesting to obtain some uncertainty
estimates for the moment-based parameter evaluations.
Moment method estimates are easier to obtain; if the
variances and covariances for parameter estimates can be
calculated (as by Evans, 1953), this would significantly
increase their value.
Though many uncertainty estimates were considered previously
(Anscombe, 1950; Shenton and Myers, 1963; Johnson {\it et
al.}, 2005), they are difficult to implement in practical
situations.

$\bullet$
3) We need to investigate the goodness-of-fit of the
distribution of earthquake numbers in various time intervals
to theoretical distributions, like the Poisson, logarithmic,
and NBD.

$\bullet$
4) Discrete distributions more general than those investigated
in this work need study.
It may be possible to establish for such distributions how
their parameters depend on earthquake catalog properties.
Then the generalized distributions can offer new insight into
earthquake number distribution.

$\bullet$
5) The lack of appropriate software hindered our analysis.
Only moment method estimates could be used for the
catalogs in all their subdivisions.
Hilbe (2007) describes many commercial and free software
packages (like SAS, S-plus, and R) that can be used for
statistical studies.
Their application would facilitate a more detailed
investigation of earthquake occurrence.
However, as few geophysicists are familiar with these
statistical codes, this may present a serious problem in
applying this software.

\subsection* {Acknowledgments
}
\label{Ackn}
Discussions with Dave Jackson and Ilya Zaliapin have been
helpful during the writing of this paper.
I am very grateful to Kathleen Jackson who edited the
manuscript.
Comments by Warner Marzocchi, by an anonymous reviewer, and
by the Associate Editor Massimo Cocco have been helpful
in revising the manuscript.
The author appreciates support from the National Science
Foundation through grant EAR-0711515, as
well as from the Southern California Earthquake Center (SCEC).
SCEC is funded by NSF Cooperative Agreement EAR-0106924 and
the U.S. Geologic Survey (USGS) Cooperative Agreement
02HQAG0008.
Publication 1304, SCEC.

\pagebreak

\centerline { {\sc References} }
\vskip 0.1in
\parskip 1pt
\parindent=1mm
\def\reference{\hangindent=22pt\hangafter=1}

\reference
Anraku, K., and T. Yanagimoto, 1990.
Estimation for the negative binomial distribution based on the
conditional likelihood,
{\sl Communications in Statistics - Simulation and
Computation}, {\bf 19}(3), 771-786.

\reference
Anscombe, F. J., 1950.
Sampling Theory of the Negative Binomial and Logarithmic
Series Distributions,
{\sl Biometrika}, {\bf 37}(3/4), 358-382.

\reference
Bartlett, M. S., 1978.
{\sl An Introduction to Stochastic Processes with Special
Reference to Methods and Applications},
Cambridge, Cambridge University Press, 3rd ed., 388 pp.

\reference
Bird, P., and Y. Y. Kagan, 2004.
Plate-tectonic analysis of shallow seismicity: apparent
boundary width, beta, corner magnitude, coupled
lithosphere thickness, and coupling in seven tectonic settings,
{\sl Bull.\ Seismol.\ Soc.\ Amer.}, {\bf 94}(6), 2380-2399
(plus electronic supplement).

\reference
Bremaud, P., and L. Massoulie, 2001.
Hawkes branching point processes without ancestors,
{\sl J. Applied Probability}, {\bf 38}(1), 122-135.

\reference
Consul, P. C., 1989.
{\sl Generalized Poisson Distributions:
Properties and Applications},
New York, Dekker, 302~pp.

\reference
Cornell, C.\ A., 1968.
Engineering seismic risk analysis,
{\sl Bull.\ Seismol.\ Soc.\ Amer.}, {\sl 58}, 1583-1606.

\reference
Dionysiou, D. D., and G. A. Papadopoulos, 1992.
Poissonian and negative binomial modelling of earthquake time
series in the Aegean area,
{\sl Phys.\ Earth Planet.\ Inter.}, {\bf 71}(3-4), 154-165.

\reference
Ebel, J. E., Bonjer, K.-P., and Oncescu, M. C., 2000.
Paleoseismicity: Seismicity evidence for past large
earthquakes,
{\sl Seismol.\ Res.\ Lett.}, {\bf 71}(2), 283-294.

\reference
Ekstr\"om, G., A. M. Dziewonski, N. N. Maternovskaya
and M. Nettles, 2005.
Global seismicity of 2003: Centroid-moment-tensor solutions
for 1087 earthquakes,
{\sl Phys.\ Earth planet.\ Inter.}, {\bf 148}(2-4), 327-351.

\reference
Enescu, B., J. Mori, M. Miyazawa, and Y. Kano, 2009.
Omori-Utsu Law $c$-Values Associated with Recent
Moderate Earthquakes in Japan,
{\sl Bull.\ Seismol.\ Soc.\ Amer.}, {\bf 99}(2A), 884-891.

\reference
Evans, D. A., 1953.
Experimental evidence concerning contagious distributions in
ecology,
{\sl Biometrika}, {\bf 40}(1-2), 186-211,

\reference
Evans, M., N. Hastings, and B. Peacock, 2000.
{\sl Statistical Distributions},
3rd ed., New York, J. Wiley, 221~pp.

\reference
Feller, W., 1968.
{\sl An Introduction to Probability Theory and its
Applications}, {\bf 1}, 3-rd ed., J. Wiley, New York, 509~pp.

\reference
Field, E. H., 2007.
Overview of the Working Group for the Development of Regional
Earthquake Likelihood Models (RELM),
{\sl Seism.\ Res.\ Lett.}, {\bf 78}(1), 7-16.

\reference
Gradshteyn, I.~S.~\& Ryzhik I. M., 1980.
{\sl Table of Integrals, Series, and Products},
Acad.\ Press, NY, pp.~1160.

\reference
Hawkes, A. G., 1971.
Point spectra of some mutually exciting point processes,
{\sl J. Roy. Statist. Soc.}, {\bf B33}, 438-443.

\reference
Hilbe, J. M., 2007.
{\sl Negative Binomial Regression},
New York, Cambridge University Press, 251~pp.

\reference
Hileman, J. A., C. R. Allen, and J. M. Nordquist, 1973.
{\sl Seismicity of the Southern California Region, 1 January
1932 to 31 December 1972}, Cal.\ Inst.\ Technology, Pasadena.

\reference
Hutton, L. K., and L. M. Jones, 1993.
Local magnitudes and apparent variations in seismicity rates in
Southern California,
{\sl Bull.\ Seismol.\ Soc.\ Am.}, {\bf 83}, 313-329.

\reference
Hutton, K., E. Hauksson, J. Clinton, J. Franck, A. Guarino, N.
Scheckel, D. Given, and A. Young, 2006.
Southern California Seismic Network update,
{\sl Seism.\ Res.\ Lett.} {\bf 77}(3), 389-395,
doi 10.1785/gssrl.77.3.389

\reference
Jackson, D.~D., and Y.~Y.~Kagan, 1999.
Testable earthquake forecasts for 1999,
{\sl Seism.\ Res.\ Lett.}, {\bf 70}(4), 393-403.

\reference
Johnson, N.\ L., A. W. Kemp, and S. Kotz, 2005.
{\sl Univariate Discrete Distributions},
3rd ed., Wiley, 646~pp.

\reference
Kagan, Y.~Y., 1973a.
A probabilistic description of the seismic regime,
{\sl Izv.\ Acad.\ Sci.\ USSR, Phys.\ Solid Earth}, 213-219,
(English translation).
Scanned versions of the Russian and English text are available
at http://eq.ess.ucla.edu/$\sim$kagan/pse\_1973\_index.html

\reference
Kagan, Y.~Y., 1973b.
Statistical methods in the study of the seismic process
(with discussion: Comments by M. S. Bartlett, A. G. Hawkes,
and J. W. Tukey),
{\sl Bull.\ Int.\ Statist.\ Inst.}, {\bf 45(3)}, 437-453.
Scanned version of text is available at
http://moho.ess.ucla.edu/$\sim$kagan/Kagan\_1973b.pdf

\reference
Kagan, Y.~Y., 1991.
Likelihood analysis of earthquake catalogues,
{\sl Geophys.\ J. Int.}, {\bf 106}(1), 135-148.

\reference
Kagan, Y.~Y., 1996.
Comment on ``The Gutenberg-Richter or characteristic
earthquake distribution, which is it?" by Steven G. Wesnousky,
{\sl Bull.\ Seismol.\ Soc.\ Amer.}, {\bf 86}(1a), 274-285.

\reference
Kagan, Y. Y., 2002.
Aftershock zone scaling,
{\sl Bull.\ Seismol.\ Soc.\ Amer.}, {\bf 92}(2), 641-655.

\reference
Kagan, Y. Y., 2003.
Accuracy of modern global earthquake catalogs,
{\sl Phys.\ Earth Planet.\ Inter.}, {\bf 135}(2-3),
173-209.

\reference
Kagan, Y. Y., 2004.
Short-term properties of earthquake catalogs and models of
earthquake source,
{\sl Bull.\ Seismol.\ Soc.\ Amer.}, {\bf 94}(4), 1207-1228.

\reference
Kagan, Y. Y., 2006.
Why does theoretical physics fail to explain and predict
earthquake occurrence?,
in: {\sl Modelling Critical and Catastrophic Phenomena in
Geoscience: A Statistical Physics Approach}, {\sl Lecture
Notes in Physics}, {\bf 705}, pp.~303-359, P. Bhattacharyya
and B. K. Chakrabarti (Eds.), Springer Verlag,
Berlin--Heidelberg.

\reference
Kagan, Y. Y., P. Bird, and D. D. Jackson, 2009.
Earthquake Patterns in Diverse Tectonic Zones of
the Globe,
accepted by {\sl Pure Appl.\ Geoph.}
({\sl Seismogenesis and Earthquake Forecasting: The Frank
Evison Volume}),
\hfil\break
http://scec.ess.ucla.edu/$\sim$ykagan/globe\_index.html .

\reference
Kagan, Y.~Y., and D.~D.~Jackson, 1991.
Long-term earthquake clustering,
{\sl Geophys.\ J. Int.}, {\bf 104}(1), 117-133.

\reference
Kagan, Y. Y., and D. D. Jackson, 2000.
Probabilistic forecasting of earthquakes,
{\sl Geophys.\ J. Int.}, {\bf 143}(2), 438-453.

\reference
Kagan, Y.~Y., and L. Knopoff, 1987.
Statistical short-term earthquake prediction,
{\sl Science}, {\bf 236}(4808), 1563-1567.

\reference
Kotz, S., N. Balakrishnan, C. Read, and B. Vidakovic, 2006.
{\sl Encyclopedia of Statistical Sciences},
2nd ed., Hoboken, N.J., Wiley-Interscience, 16 vols.

\reference
Lombardi, A. M. and W. Marzocchi, 2007.
Evidence of clustering and nonstationarity in the time
distribution of large worldwide earthquakes, {\sl J. Geophys.\
Res.}, {\bf 112}, B02303, doi:10.1029/2006JB004568.

\reference
Ogata, Y., 1988.
Statistical models for earthquake occurrence and residual
analysis for point processes,
{\sl J.\ Amer.\ Statist.\ Assoc.}, {\bf 83}, 9-27.

\reference
Ogata, Y., 1998.
Space-time point-process models for earthquake occurrences,
{\sl Ann.\ Inst.\ Statist.\ Math.}, {\bf 50}(2), 379-402.

\reference
Patil, G. P., 1962.
Some methods of estimation for logarithmic series
distribution,
{\sl Biometrics}, {\bf 18}(1), 68-75.

\reference
Patil, G. P., and J. K. Wani, 1965.
Maximum likelihood estimation for the complete and truncated
logarithmic series distributions,
{\sl Sankhya}, {\bf 27A}(2/4), 281-292.

\reference
{\sl Preliminary determinations of epicenters (PDE)}, 2008.
U.S.\ Geological Survey,
U.S.\ Dep.\ of Inter., Natl.\ Earthquake Inf.\ Cent.,
http://neic.usgs.gov/neis/epic/epic.html and
http://neic.usgs.gov/neis/epic/code\_catalog.html .

\reference
Schorlemmer, D., and M. C. Gerstenberger, 2007.
RELM testing Center,
{\sl Seism.\ Res.\ Lett.}, {\bf 78}(1), 30-35.

\reference
Schorlemmer, D., M. C. Gerstenberger, S. Wiemer, D. D.
Jackson, and D. A. Rhoades, 2007.
Earthquake likelihood model testing,
{\sl Seism.\ Res.\ Lett.}, {\bf 78}(1), 17-29.

\reference
Schorlemmer, D., J. D. Zechar, M. Werner, D. D. Jackson, E. H.
Field, T. H. Jordan, and the RELM Working Group, 2009.
First results of the Regional Earthquake Likelihood Models
Experiment,
{\sl Pure and Applied Geophysics}, accepted.

\reference
Shenton, L. R., and R. Myers, 1963,
Comments on estimation for the negative binomial distribution,
in {\sl Classical and Contagious Discrete Distributions}, G.
P. Patil, Ed., Stat.\ Publ.\ Soc., Calcutta, pp.~241-262.

\reference
Shlien, S., and Toks\"oz, M.N., 1970.
A clustering model for earthquake occurrences.
{\sl Bull.\ Seismol.\ Soc.\ Amer.}, {\bf 60}(6), 1765-1788.

\reference
Sornette, D., and M.J. Werner, 2005.
Apparent clustering and apparent background
earthquakes biased by undetected seismicity,
{\sl J. Geophys.\ Res.}, {\bf 110}(9), B09303,
doi:10.1029/2005JB003621 .

\reference
Tripathi, R. C., 2006.
Negative binomial distribution,
in: {\sl Encyclopedia of Statistical Sciences}, Kotz, S., N.
Balakrishnan, C. Read, and B. Vidakovic, Eds., 2nd ed.,
Hoboken, N.J., Wiley-Interscience, vol. {\bf 8}, pp.~5413-20.

\reference
Vere-Jones, D., 1970.
Stochastic models for earthquake occurrence (with discussion),
{\sl J.\ Roy.\ Stat.\ Soc.}, {\bf B32}(1), 1-62.

\reference
Wilks, S.\ S.,
{\sl Mathematical Statistics},
John Wiley, New York, 1962, 644~pp.

\reference
Zaliapin, I. V., Y. Y. Kagan, and F. Schoenberg, 2005.
Approximating the distribution of Pareto sums,
{\sl Pure Appl.\ Geoph.}, {\bf 162}(6-7), 1187-1228.

\reference
Zhuang, J. C., Y. Ogata, and D. Vere-Jones, 2004.
Analyzing earthquake clustering features by using stochastic
reconstruction,
{\sl J. Geophys.\ Res.}, {\bf 109}(B5), Art.\ No.\ B05301.

\newpage


\begin{table}
\caption{Values of NBD parameters for various subdivisions of
the 1969-2007 PDE catalog
}
\vspace{10pt}
\label{Table1}
\begin{tabular}{lcrrrcccrr}
\hline
& & & & & & & & & \\[-25pt]
\multicolumn{1}{c}{Subd}&
\multicolumn{1}{c}{$m_c$}&
\multicolumn{1}{c}{$n$}&
\multicolumn{1}{c}{$N$}&
\multicolumn{1}{c}{$\lambda$}&
\multicolumn{1}{c}{$\alpha$}&
\multicolumn{1}{c}{$a$}&
\multicolumn{1}{c}{$\theta$}&
\multicolumn{1}{c}{$\tau$}&
\multicolumn{1}{c}{$\Delta T$}
\\[2pt]
\hline
\multicolumn{1}{c}{1}&
\multicolumn{1}{c}{2}&
\multicolumn{1}{c}{3}&
\multicolumn{1}{c}{4}&
\multicolumn{1}{c}{5}&
\multicolumn{1}{c}{6}&
\multicolumn{1}{c}{7}&
\multicolumn{1}{c}{8}&
\multicolumn{1}{c}{9}&
\multicolumn{1}{c}{10}
\\[2pt]
\hline
& & & & & & & & & \\[-15pt]
G & 7.0 & 459 & 39 & 11.8 & 0.0089 & 0.1044 & 0.9055 & 112.7 & 365.2 \\
G & 6.5 & 1359 & 39 & 34.9 & 0.0207 & 0.7212 & 0.5810 & 48.32 & 365.2 \\
G & 6.0 & 3900 & 39 & 100.0 & 0.0284 & 2.835 & 0.2607 & 35.27 & 365.2 \\
G & 5.5 & 13553 & 39 & 347.5 & 0.0395 & 13.72 & 0.0679 & 25.33 & 365.2 \\
G & 5.0 & 47107 & 39 & 1207.9 & 0.0335 & 40.504 & 0.0241 & 29.82 & 365.2 \\
& & & & & & & & & \\[-15pt]
G & 5.0 & 47107 & 5 & 9421.4 & 0.0118 & 111.06 & 0.0089 & 84.83 & 2848.8 \\
G & 5.0 & 47107 & 10 & 4710.7 & 0.0269 & 126.52 & 0.0078 & 37.24 & 1424.4 \\
G & 5.0 & 47107 & 20 & 2355.4 & 0.0309 & 72.72 & 0.0136 & 32.39 & 712.2 \\
G & 5.0 & 47107 & 39 & 1207.9 & 0.0335 & 40.50 & 0.0241 & 29.82 & 365.2 \\
G & 5.0 & 47107 & 50 & 942.1 & 0.0422 & 39.78 & 0.0245 & 23.68 & 284.9 \\
G & 5.0 & 47107 & 100 & 471.1 & 0.0541 & 25.50 & 0.0377 & 18.47 & 142.4 \\
G & 5.0 & 47107 & 200 & 235.5 & 0.0670 & 15.78 & 0.0596 & 14.92 & 71.2 \\
G & 5.0 & 47107 & 500 & 94.2 & 0.1137 & 10.72 & 0.0854 & 8.792 & 28.5 \\
G & 5.0 & 47107 & 1000 & 47.1 & 0.1620 & 7.632 & 0.1158 & 6.172 & 14.2 \\
& & & & & & & & & \\[-15pt]
0 & 5.0 & 2225 & 39 & 57.1 & 0.1686 & 9.621 & 0.0942 & 5.930 & 365.2 \\
1 & 5.0 & 7740 & 39 & 198.5 & 0.0645 & 12.80 & 0.0725 & 15.50 & 365.2 \\
2 & 5.0 & 3457 & 39 & 88.6 & 0.0722 & 6.400 & 0.1351 & 13.85 & 365.2 \\
3 & 5.0 & 3010 & 39 & 77.2 & 0.1475 & 11.38 & 0.0808 & 6.780 & 365.2 \\
4 & 5.0 & 30675 & 39 & 786.5 & 0.0329 & 25.85 & 0.0372 & 30.43 & 365.2 \\
& & & & & & & & & \\[-15pt]
\hline
\end{tabular}

\bigskip
In column 1: G means that the global catalog is used, 0 --
plate interior, 1 -- Active continent, 2 -- Slow ridge, 3 --
Fast ridge, 4 -- Trench (subduction zones), see Kagan {\it et
al.}, 2009; $n$ is the number of earthquakes,
$N$ is the number of time intervals, $\Delta T$ -- interval
duration in days.

\hfil\break
\vspace{5pt}
\end{table}

\newpage

\begin{table}
\caption{Values of NBD parameters for various subdivisions of
the 1977-2007 CMT catalog
}
\vspace{10pt}
\label{Table2}
\begin{tabular}{lcrrrrrrrr}
\hline
& & & & & & & & & \\[-25pt]
\multicolumn{1}{c}{Subd}&
\multicolumn{1}{c}{$m_c$}&
\multicolumn{1}{c}{$n$}&
\multicolumn{1}{c}{$N$}&
\multicolumn{1}{c}{$\lambda$}&
\multicolumn{1}{c}{$\alpha$}&
\multicolumn{1}{c}{$a$}&
\multicolumn{1}{c}{$\theta$}&
\multicolumn{1}{c}{$\tau$}&
\multicolumn{1}{c}{$\Delta T$}
\\[2pt]
\hline
\multicolumn{1}{c}{1}&
\multicolumn{1}{c}{2}&
\multicolumn{1}{c}{3}&
\multicolumn{1}{c}{4}&
\multicolumn{1}{c}{5}&
\multicolumn{1}{c}{6}&
\multicolumn{1}{c}{7}&
\multicolumn{1}{c}{8}&
\multicolumn{1}{c}{9}&
\multicolumn{1}{c}{10}
\\[2pt]
\hline
& & & & & & & & & \\[-15pt]
G & 7.0 & 307 & 31 & 9.903 & $-0.0211$ & $-0.2094$ & 1.2649 & $-47.29$ & 365.2 \\
G & 6.5 & 1015 & 31 & 32.742 & 0.0159 & 0.5211 & 0.6574 & 62.83 & 365.2 \\
G & 6.0 & 3343 & 31 & 107.839 & 0.0181 & 1.956 & 0.3384 & 55.15 & 365.2 \\
G & 5.8 & 5276 & 31 & 170.194 & 0.0138 & 2.356 & 0.2980 & 72.25 & 365.2 \\
& & & & & & & & & \\[-15pt]
G & 5.8 & 5276 & 5 & 1055.200 & 0.0026 & 2.706 & 0.2698 & 389.9 & 2264.4 \\
G & 5.8 & 5276 & 10 & 527.600 & 0.0076 & 3.996 & 0.2001 & 132.0 & 1132.2 \\
G & 5.8 & 5276 & 20 & 263.800 & 0.0107 & 2.821 & 0.2617 & 93.52 & 566.1 \\
G & 5.8 & 5276 & 31 & 170.194 & 0.0138 & 2.356 & 0.2980 & 72.25 & 365.2 \\
G & 5.8 & 5276 & 50 & 105.520 & 0.0220 & 2.320 & 0.3012 & 45.48 & 226.4 \\
G & 5.8 & 5276 & 100 & 52.760 & 0.0268 & 1.411 & 0.4147 & 37.38 & 113.2 \\
G & 5.8 & 5276 & 200 & 26.380 & 0.0417 & 1.100 & 0.4763 & 23.99 & 56.6 \\
G & 5.8 & 5276 & 500 & 10.552 & 0.0795 & 0.8392 & 0.5437 & 12.57 & 22.6 \\
G & 5.8 & 5276 & 1000 & 5.276 & 0.1079 & 0.5693 & 0.6372 & 9.267 & 11.3 \\
& & & & & & & & & \\[-15pt]
0 & 5.8 & 172 & 31 & 5.548 & 0.0604 & 0.3353 & 0.7489 & 16.55 & 365.2 \\
1 & 5.8 & 723 & 31 & 23.323 & 0.0357 & 0.8323 & 0.5458 & 28.02 & 365.2 \\
2 & 5.8 & 336 & 31 & 10.839 & 0.0528 & 0.5720 & 0.6361 & 18.95 & 365.2 \\
3 & 5.8 & 537 & 31 & 17.323 & 0.0234 & 0.4055 & 0.7115 & 42.72 & 365.2 \\
4 & 5.8 & 3508 & 31 & 113.161 & 0.0268 & 3.033 & 0.2480 & 37.32 & 365.2 \\
& & & & & & & & & \\[-15pt]
\hline
\end{tabular}

\bigskip
In column 1: G means that the global catalog is used, 0 --
plate interior, 1 -- Active continent, 2 -- Slow ridge, 3 --
Fast ridge, 4 -- Trench (subduction zones), see Kagan {\it et
al.}, 2009; $n$ is the number of earthquakes,
$N$ is the number of time intervals, $\Delta T$ -- interval
duration in days.

\hfil\break
\vspace{5pt}
\end{table}

\newpage

\begin{table}
\caption{Values of NBD parameters for various subdivisions of
the 1932-2001 CIT catalog
}
\vspace{10pt}
\label{Table3}
\begin{tabular}{lcrrrcccrr}
\hline
& & & & & & & & & \\[-25pt]
\multicolumn{1}{c}{Subd}&
\multicolumn{1}{c}{$m_c$}&
\multicolumn{1}{c}{$n$}&
\multicolumn{1}{c}{$N$}&
\multicolumn{1}{c}{$\lambda$}&
\multicolumn{1}{c}{$\alpha$}&
\multicolumn{1}{c}{$a$}&
\multicolumn{1}{c}{$\theta$}&
\multicolumn{1}{c}{$\tau$}&
\multicolumn{1}{c}{$\Delta T$}
\\[2pt]
\hline
\multicolumn{1}{c}{1}&
\multicolumn{1}{c}{2}&
\multicolumn{1}{c}{3}&
\multicolumn{1}{c}{4}&
\multicolumn{1}{c}{5}&
\multicolumn{1}{c}{6}&
\multicolumn{1}{c}{7}&
\multicolumn{1}{c}{8}&
\multicolumn{1}{c}{9}&
\multicolumn{1}{c}{10}
\\[2pt]
\hline
& & & & & & & & & \\[-15pt]
G & 6.0 & 25 & 70 & 0.357 & 2.1360 & 0.7629 & 0.56726 & 0.4682 & 365.2 \\
G & 5.0 & 226 & 70 & 3.229 & 1.4422 & 4.6564 & 0.17679 & 0.6934 & 365.2 \\
G & 4.0 & 2274 & 70 & 32.49 & 1.2976 & 42.154 & 0.02317 & 0.7706 & 365.2 \\
G & 3.0 & 17393 & 70 & 248.5 & 0.9184 & 228.20 & 0.00436 & 1.0888 & 365.2 \\
G & 2.0 & 5391 & 13 & 414.7 & 1.2958 & 537.36 & 0.00186 & 0.7717 & 365.2 \\
& & & & & & & & & \\[-15pt]
G & 3.0 & 17393 & 5 & 3478.6 & 0.1062 & 369.46 & 0.00270 & 9.4155 & 5113.6 \\
G & 3.0 & 17393 & 10 & 1739.3 & 0.1569 & 272.87 & 0.00365 & 6.3740 & 2556.8 \\
G & 3.0 & 17393 & 20 & 869.7 & 0.3389 & 294.73 & 0.00338 & 2.9507 & 1278.4 \\
G & 3.0 & 17393 & 50 & 347.9 & 0.6514 & 226.58 & 0.00439 & 1.5353 & 511.4 \\
G & 3.0 & 17393 & 70 & 248.5 & 0.9184 & 228.20 & 0.00436 & 1.0888 & 365.2 \\
G & 3.0 & 17393 & 100 & 173.9 & 1.1803 & 205.28 & 0.00485 & 0.8473 & 255.7 \\
G & 3.0 & 17393 & 200 & 86.97 & 1.7691 & 153.85 & 0.00646 & 0.5652 & 127.8 \\
G & 3.0 & 17393 & 500 & 34.79 & 3.9832 & 138.56 & 0.00717 & 0.2511 & 51.1 \\
G & 3.0 & 17393 & 1000 & 17.39 & 6.5226 & 113.45 & 0.00874 & 0.1533 & 25.6 \\
& & & & & & & & & \\[-15pt]
SE & 3.0 & 8281 & 70 & 118.3 & 1.7967 & 212.55 & 0.00468 & 0.5566 & 365.2 \\
NE & 3.0 & 2839 & 70 & 40.56 & 4.2550 & 172.57 & 0.00576 & 0.2350 & 365.2 \\
SW & 3.0 & 2622 & 70 & 37.46 & 3.7734 & 141.34 & 0.00703 & 0.2650 & 365.2 \\
NW & 3.0 & 3651 & 70 & 52.16 & 1.7630 & 91.953 & 0.01076 & 0.5672 & 365.2 \\
& & & & & & & & & \\[-15pt]
\hline
\end{tabular}

\bigskip
In column 1: G means that the whole CIT catalog is used, SE --
south-east part of southern California, NE -- north-east, SW
-- south-west, NW -- north-west;
$n$ is the number of earthquakes,
$N$ is the number of time intervals, $\Delta T$ -- interval
duration in days.

\hfil\break
\vspace{5pt}
\end{table}

\newpage

\clearpage

\parindent=0mm

\begin{figure}
\begin{center}
\includegraphics[width=0.75\textwidth]{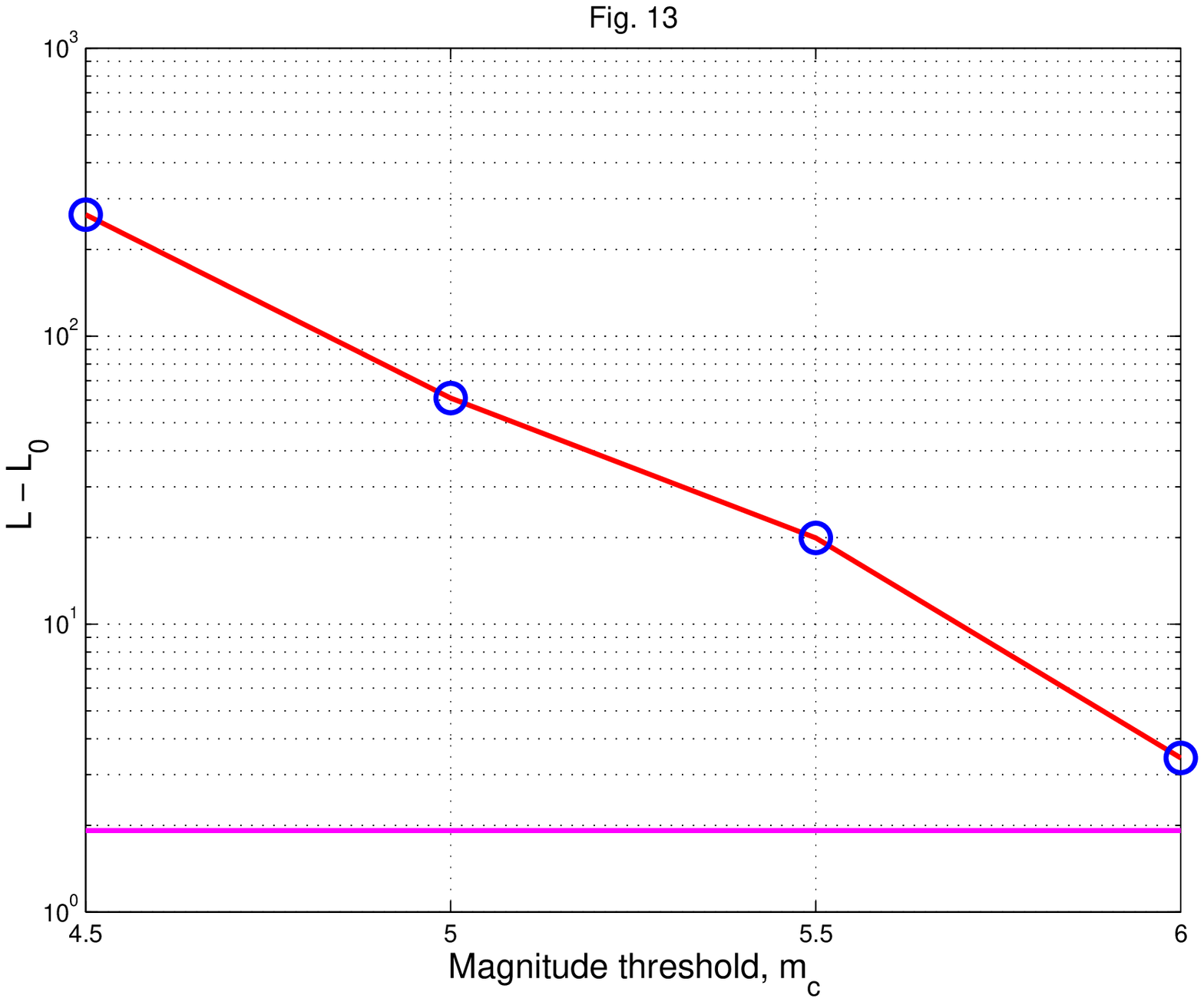}
\caption{\label{fig13}
}
\end{center}
Dependence of the log-likelihood difference for the NBD and
Poisson models of earthquake occurrence on the threshold
magnitude.
The CIT catalog 1932-2001 is used, annual event numbers are
analyzed.
The magenta line corresponds to $\ell - \ell_0 = 1.92$; for a
higher log-likelihood difference level the Poisson hypothesis
should be rejected at the 95\% confidence limit.
\end{figure}

\begin{figure}
\begin{center}
\includegraphics[width=0.75\textwidth]{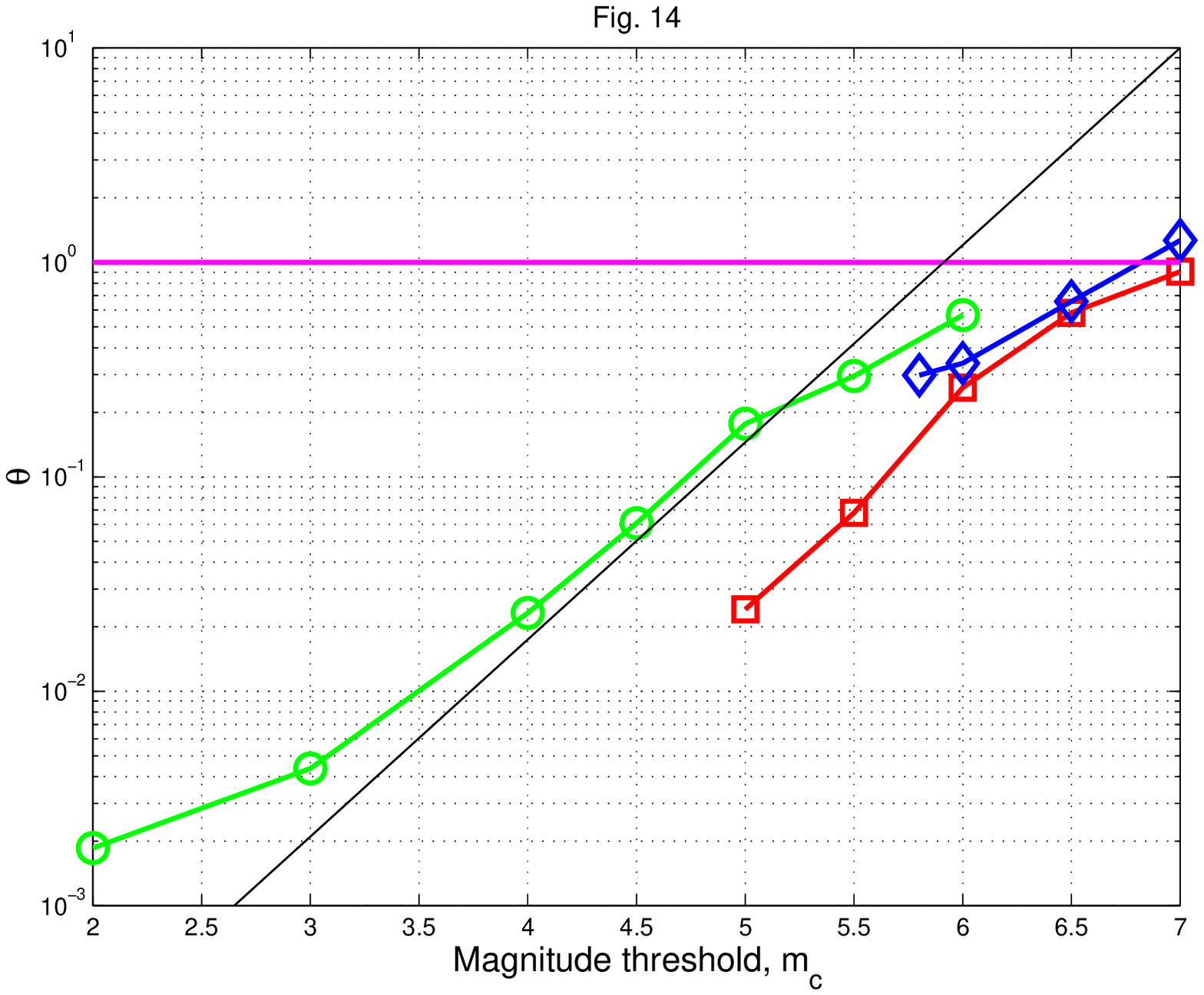}
\caption{\label{fig14}
}
\end{center}
Dependence of the $\theta$-value for the NBD model
(\ref{NBD_Eq15}) of earthquake occurrence on the threshold
magnitude.
Three catalogs are used:
the green curve is for the CIT catalog 1932-2001,
the red curve is for the PDE catalog 1969-2007, and
the blue curve is for the CMT catalog 1977-2007.
The magenta line is $\theta = 1.0$, corresponding to the
Poisson occurrence.
Thin black line corresponds to $\theta \propto 10^{1.5 \,
m \, \beta}$ (see Eq.~\ref{NBD_Eq04}).
\end{figure}

\begin{figure}
\begin{center}
\includegraphics[width=0.75\textwidth]{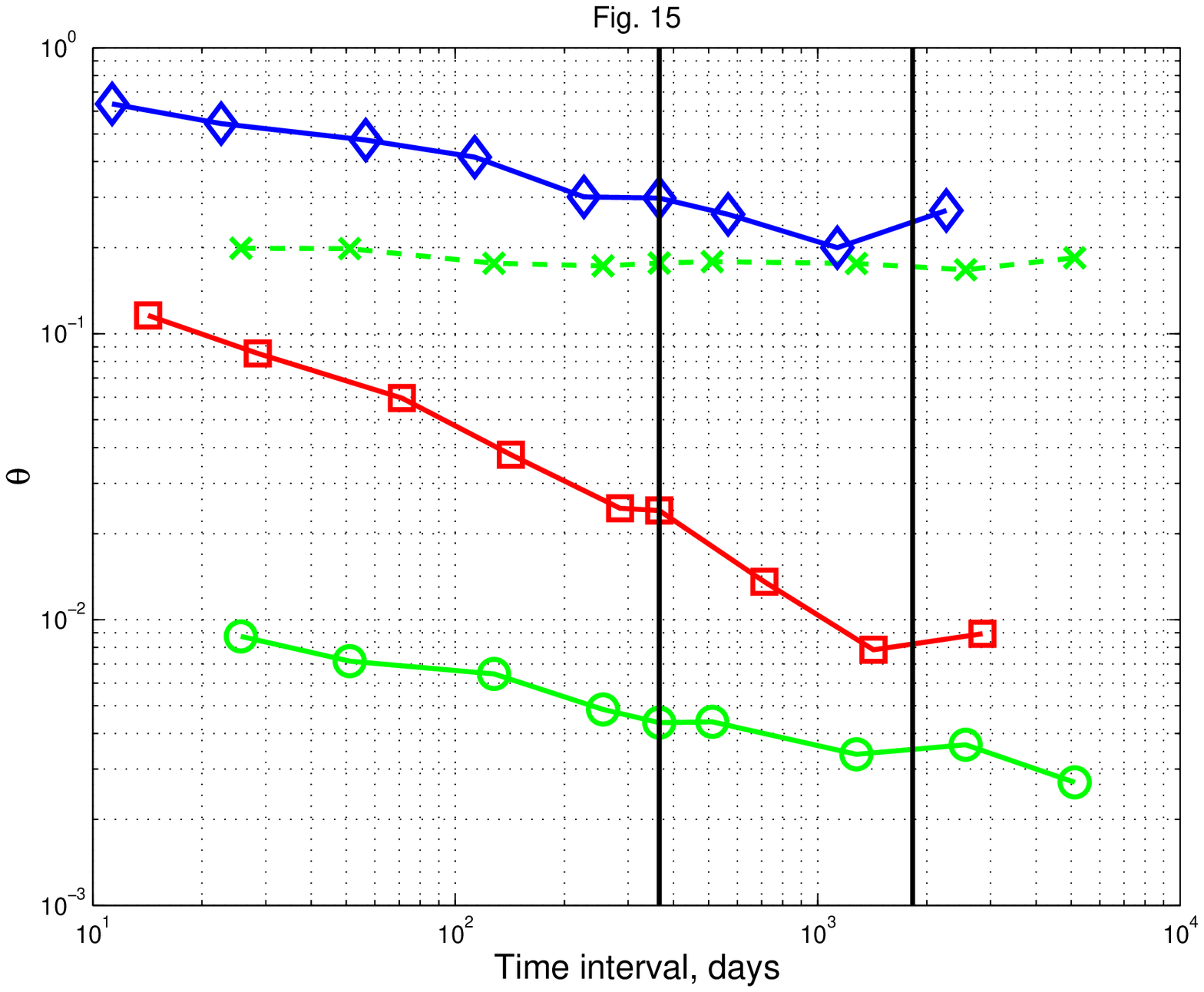}
\caption{\label{fig15}
}
\end{center}
Dependence of the $\theta$-value for the NBD model
(\ref{NBD_Eq15}) of earthquake occurrence on time interval
duration ($\Delta T$).
Three catalogs are used:
the green curve is for the CIT catalog 1932-2001 (circles are
for $m3$ earthquakes and crosses for $m5$);
the red curve is for the PDE catalog 1969-2007, and
the blue curve is for the CMT catalog 1977-2007.
Two black vertical lines show 1-year and 5-year intervals.
\end{figure}


\end{document}